\documentclass{aa}

\usepackage{graphicx}
\usepackage{txfonts}
\usepackage{natbib}
\bibpunct{(}{)}{;}{a}{}{,}

\def\harrow{\mathrel{\hbox{\rlap{\hbox{\raise4pt\hbox{${\rm +H}$}}}\hbox{$\longrightarrow$}}}}
\def\oarrow{\mathrel{\hbox{\rlap{\hbox{\raise4pt\hbox{${\rm +O}$}}}\hbox{$\longrightarrow$}}}}

\def\lesssim{\mathrel{\hbox{\rlap{\hbox{\lower4pt\hbox{$\sim$}}}\hbox{$<$}}}}
\def\gtsim{\mathrel{\hbox{\rlap{\hbox{\lower4pt\hbox{$\sim$}}}\hbox{$>$}}}}

\begin{document}

\title{Physical structure and water line spectrum predictions of the
  Intermediate Mass protostar OMC2-FIR4} 
\author{N.Crimier\inst{1} \and C. Ceccarelli\inst{1} \and
  B. Lefloch\inst{1} \and A. Faure\inst{1} }

\offprints{N.Crimier}

\institute{
$^1$Laboratoire d'Astrophysique de l'Observatoire de Grenoble, 
BP 53, 38041 Grenoble, Cedex 9, France. \\
\email{ncrimier,ceccarel,lefloch,afaure@obs.ujf-grenoble.fr}
}

\date{Accepted}

\abstract
{}
{ Intermediate Mass (IM) stars are an important component of our
  Galaxy, as they significantly contribute to the interstellar FUV
  field and, consequently, play an important role in the energy
  balance of the ISM. Despite their importance, very little is known
  about their formation process and only a few studies have been
  devoted to characterize the first phases in the evolution of
  intermediate mass protostars. Here we consider in great detail the
  case of the brightest and closest known young IM protostar: FIR4 in
  the OMC2 component of the Orion molecular cloud complex.}
{ We analyzed the available continuum emission (maps and SED) through
  one-dimensional dust radiative transfer calculations. We ran large
  grids of models to find the envelope model that best fits the
  data. The derived dust density and temperature profiles have been
  then used to compute the gas temperature profile, equating gas
  cooling and heating terms across the envelope. Last, we computed the
  water line spectrum for various possible values of water abundance.}
{ The luminosity of FIR4 has been reevaluated to 1000 L$_\odot$, making
  FIR4 definitively an Intermediate Mass protostar. The envelope
  surrounding FIR4 has a relatively shallow density power law index,
  $\sim0.6$. The most surprising result is that the gas and dust are
  thermally decoupled in the interior of the envelope, where the dust
  ices sublimate at 100 K. This has important consequences in the
  interpretation of the line data. We provide the predictions for the
  water spectrum, and discuss in detail the lines which will be
  observed by the Herschel Space Observatory .}
{}

\titlerunning{the physical structure of OMC2-FIR4}
\authorrunning{N.Crimier et al.}

\maketitle

\section{Introduction}\label{sec:introduction}
Intermediate mass (IM) stars, namely stars whose mass is in the 2 to 8
M$_\odot$ range, are crucial in studies of star formation because they
constitute the link between low- and high-mass stars
\citep{DiF97,Man97,Man00}, and, therefore, can help to understand if
and how much different are the processes at work in the two ends.  On
the one hand, low mass stars are can be formed isolated or in loose
groups of few objects per cubic parsec \citep{Gom93}, while high-mass
stars are usually found to form in tight clusters
\citep[e.g.][]{Hil98}. IM stars, on the other hand, are also found in
clusters \citep[e.g.][]{Tes98,Ner07,Fue07}, with a smooth transition
towards the low mass star, loose cluster regime for star masses around
3.5 M$_\odot$ \citep{Tes99}. \citet{Tes99} also concluded that IM
stars mark the transition from low density aggregates of $\lesssim$ 10
stars per cubic parsec of T Tauri stars to dense clusters of $\gtrsim$
10$^3$ stars per cubic parsec associated with early-type stars. In
agreement with the different observed environments, several authors
have proposed that high mass stars are formed by coalescence of lower
mass stars, whereas other authors favor the ``monolithic''
formation (see for example the recent review by \citet{Beu07}). In
this context, the IM stars study can greatly help the debate.  Indeed,
due to their intermediate position, the study of IM protostars will
provide crucial information on the transition between the two
formation regimes as well as on the limits of the low mass and high
mass formation scenarios.  Finally, IM stars are among the dominant
sources of the Inter-Stellar FUV field \citep[e.g.][]{Hab68,Gon75},
which regulates the phases of the ISM in the Galaxy, and, in turns,
the overall Galaxy star formation process and history. Despite the
far-reaching importance of IM stars, very little is known about the
formation and first evolutionary stages of these stars. The situation
is so bad that to date we do not have a satisfying sample of Class 0
IM objects, namely objects representing the first phases of stellar
formation, where the protostar is embedded in its envelope and its
luminosity is dominated by the accretion luminosity, nor a systematic
study of their physical structure, as it is the case for low mass
Class 0 sources \citep[e.g.][]{Cec07,Dif07}.  This article is the
first of a series that aims to fill this gap in our knowledge.

In this context, the Orion Molecular Cloud 2 (OMC2), the closest known
region where high to low mass star formation is going on, represents a
precious laboratory for these studies. Observed first by
\citet{Gat74}, OMC2 is located 15' ($\sim$ 2 pc) North of the Orion
nebula. It has a filamentary structure, elongated in the direction
north-south, with active star formation concentrated in the central
and densest region, shielded from the UV radiation from newly formed
OB stars \citep{Joh90}. The mass of the cloud amounts to about 1500
M$_\odot$ \citep{Mez90}. Several extensive studies have shown that
OMC2 is a rich star forming region, which harbors several young
protostars, including several Class 0 candidates
\citep{Ali95,Chi97,Lis98,Johnstone99,Rei99}.  Observations of
molecular lines have revealed several outflows emanating from the
young protostars in the region. Many studies have focused on the
outflows \citep[e.g.][]{Wil03} and their impact on the cloud
\citep{Aso00,Wu05}. Only few of these studies, in contrast, have
addressed the problem of the chemical structure of the forming stars
in OMC2 \citep{Johnstone03}.

Among the several protostars in OMC2, FIR4 stands out as the brightest
submillimeter source \citep{Mez90}. Located almost at the center of
the cloud, FIR4 is also a bright IRAS source and a VLA radio source
\citep{Rei99}. All these characteristics led \citet{Rei99} to define
FIR4 ``a bona fide Class 0 source''. The FIR4 integrated luminosity
was estimated to be about 400 L$_\odot$ and the envelope mass is about
35 M$_\odot$. Such values led to identify FIR4 as an {\it intermediate
  mass protostar} \citep{Johnstone03}. Because of its vicinity and its
relatively bright molecular lines, FIR4 is an ideal source for a
detailed study of the physical and chemical structure of an IM
protostar. Existing dust continuum and molecular line observations
point to an envelope with at least two components: a warm component
with an average temperature of about 40 K and a colder component at
about 15 K \citep{Mez90,Johnstone03}.  \citet{Jor06} modeled the 850
$\mu$m SCUBA map towards this source to reconstruct its temperature
and density profiles. Based on the observed CO and H$_2$CO millimeter
line emission, \citet{Jor06} concluded that the FIR4 envelope is
illuminated by an external FUV field amounting to $1\times10^{4}$
times the Interstellar FUV field. However, their interpretation
suffers of some contradictions emphasized by the same authors. For
example, such an intense FUV field would heat up the whole envelope to
a temperature larger than 25 K, the CO freezing temperature \citep{Obe05}, in contradiction with the measured average CO
abundance, ten times lower than the canonical value, which would
rather testify for a large CO-frozen region \citep{Jor06}. In
addition, the maps of the fine structure lines of the O and C$^+$
atoms together with the CO 1-0 line led \citet{Her97} to conclude that
the OMC2 region is illuminated by a FUV field 500 times the
Interstellar field.

Given this puzzling situation, we decided to derive again the
temperature and density profiles of FIR4 by taking into account more
data than those considered by \citet{Jor06} (\S
\ref{sec:dust-dens-temp}). Using the derived dust temperature and
density profiles, we then computed the gas temperature profile, by
equating the heating and cooling terms across the envelope (\S
\ref{sec:gas-temp-prof}). As shown by several authors
\citep[e.g.][]{Cec96,Dot97}, the gas cooling in protostellar envelopes
is dominated by the emission from the rotational lines of CO and, more
important, H$_2$O together with the fine structure lines of
OI. Actually, water is a key molecule in the gas thermal balance for
two reasons. First, in the warm regions where the grain mantles
sublimate, it is the most abundant molecule; second, given its
relatively large dipole moment, water is a very powerful line emitter
and, consequently, gas coolant.  Given its major role in the
prediction of the gas temperature profile, we discuss the dependence
of the derived gas temperature on the assumed water abundance profile,
which is poorly known. Not surprising, FIR4 is in fact one of the few
sources where the full spectrum between 50 and 2000 GHz is planned to
be observed at high spectral resolution with the Heterodyne Instrument
for the Far Infrared (HIFI) on board Herschel
(http://herschel.esac.esa.int/), to be launched in 2009. HSO, and
specifically the high resolution interferometer HIFI, will allow to
observe the water lines in the 500 to 2000 GHz range with
unprecedented spectral and spatial resolution. Motivated by the
Herschel mission, we report the predicted water line spectrum for the
different assumed water abundance profiles, and discuss the
observability by HIFI and PACS (\S
\ref{sec:predicted-water-line}).  Section \ref{sec:conclusion}
concludes the article.

\section{Dust density and temperature profiles}\label{sec:dust-dens-temp}
In this section, we derive the dust density and temperature profiles
by modelling the 350, 450 and 850 $\mu$m maps of the region, plus the
Spectral Energy Distribution (SED) from the millimeter to the
Mid-Infrared (MIR) wavelength range. We first describe the
observations we used in our analysis (\S \ref{sec:cont-emiss-data})
and then the modeling (\S \ref{sec:cont-emiss-dusty}) and the result
of the modeling (\S \ref{sec:dust-results}).

\subsection{Continuum emission: observational data}\label{sec:cont-emiss-data}
In our analysis, we use the maps of the continuum emission at 850, 450
and 350 $\mu m$ obtained at JCMT and CSO respectively. In addition, we
take into account the Spectral Energy Distribution (SED) of FIR4 from
24 to 850 $\mu$m obtained considering also the IRAS and Spitzer
observations.

\begin{figure*} \centering
\rotatebox{270}{\includegraphics[width=10cm]{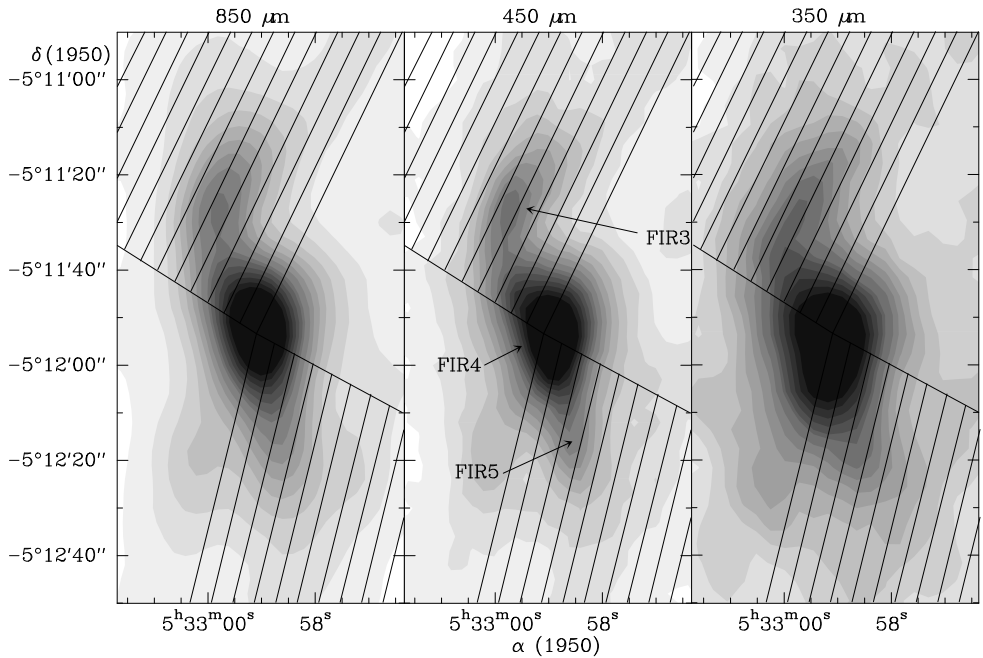}}
\caption{Continuum emission maps around OMC2-FIR4 at 850 $\mu$m (left
  panel), 450 $\mu$m (middle panel) and 350 $\mu$m (right
  panel). The contours mark the continuum flux from 5 \% to 75 \% of
  the peak emission by steps of 5 \%. The hatched regions have been
  excluded when computing the brightness profile of the FIR4 envelope
  (see text). The position of the three protostars in the regions,
  FIR3, FIR4 and FIR 5 are marked in the central panel
  figure. \label{maps}}
\end{figure*}

\noindent
{\it a) 850, 450 and 350 $\mu m$ maps}\\
We retrieved the 450 and 850 $\mu$m maps obtained by \citet{Johnstone99}
 at the 15 m James Clerk Maxwell Telescope (JCMT) with the
focal-plane instrument SCUBA (Submillimeter Common-User Bolometer
Array). The spatial resolution of the maps is 7.5$"$ and 14.8$"$ at
450 and 850 $\mu$m respectively. The calibration uncertainty and
noise levels are estimated by those authors $\lesssim$ 10\% and 0.04
Jy beam$^{-1}$ at 850 $\mu$m and $\lesssim$ 30\% and 0.3 Jy
beam$^{-1}$ at 450 $\mu$m, respectively. The 350 $\mu$m map was
obtained by \citet{Lis98} at the 10.4 m telescope of the Caltech
Submillimeter Observatory (CSO). The instrument used was the bolometer
camera SHARC. The resolution of the map is 12$"$. The calibration
uncertainty has been evaluated $\sim$ 25\%-30\%. The three maps are
reported in Fig. \ref{maps}.
They show the envelope surrounding/forming FIR4 which extends for
about 20$"$, but also the presence of two sources: FIR3, 25$"$ North,
and FIR5, 25$"$ South. To evaluate the continuum brightness profile of
the FIR4 envelope, we averaged the continuum flux over ring at the
same distance from the FIR4 center, excluding the regions contaminated
by the presence of FIR3 and FIR5 (dashed regions in
Fig. \ref{maps}). The resulting brightness profiles are shown in
Fig. \ref{fit}. Note that in the analysis of the envelope emission (\S
\ref{sec:cont-emiss-dusty}) we subtracted the cloud contribution,
estimated to be $\sim$ 0.001, $\sim$ 0.03 and $\sim$ 0.05 Jy
arcsec$^{-2}$ at 850, 450 and 350 $\mu$m respectively. Furthermore, in
order to take into account that the SCUBA and SHARC maps were obtained
with the chop throw of 65$"$ and 90-120$"$ respectively, we only
considered the inner 60$"$ in our analysis.

\noindent
{\it b) SED}\\
The SED points at 850, 450 and 350 $\mu$m, shown in Fig. \ref{fit},
were obtained integrating the continuum emission over the envelope. We
attributed an uncertainty of $\sim$ 30 \% to them to account for the
uncertainty in the envelope size. We also considered the IRAS fluxes
at 60 and 100 $\mu$m, respectively, extracted from the IRAS maps at
these wavelengths.  The evaluation of the fluxes was done using the
method previously employed for the maps at 850, 450 and 350 $\mu$m,
namely excluding the same regions (dashed regions in Fig. \ref{maps})
to limit the contamination by FIR3 and FIR5 and integrating over the
rings.  We also subtracted the cloud contribution, estimated to be
$\sim$ 0.06 and $\sim$ 0.07 Jy arcsec$^{-2}$ at 60 and 100 $\mu m$
respectively. To account for the possible contamination of FIR3 and
FIR5 due to the large beam of IRAS and the non-sphericity of the
source, we took an uncertainty of 50 \% on the fluxes.  Finally, we
also considered the integrated flux at 24 $\mu$m extracted from the
Spitzer Space Telescope's Multiband Imaging Photometer (MIPS) maps. To
this end, we retrieved the observations from the Spitzer Science
archive ({\it http://ssc.spitzer.caltech.edu/archanaly/}). The
observations were obtained the 6th October 2006 as part of the Program
``Infrared Properties of Edge-on Young Stellar Object Disks'' (AOR:
30765, PI: Karl Stapelfeldt).  The data reduction was performed using
the pipeline S16.0.1. The flux, ($5.0\pm2.5$ Jy), in Fig. \ref{fit}
was obtained by integration over a 15$"$ radius.

\subsection{Continuum emission: modeling}\label{sec:cont-emiss-dusty}
To derive the dust physical structure, namely the dust temperature and
density profiles, we used the 1D radiative transfer code DUSTY
\citep{Ive97}.  Briefly, giving as input the temperature of the
central object and a dust density profile, DUSTY computes
self-consistently the dust temperature profile and the dust
emission. The comparison between the computed 350, 450, 850 $\mu$m
brightness profiles (namely the brightness versus the distance from
the center of the envelope) and SED with the observed profiles and SED
(see previous paragraph) makes it possible to constrain the density
profile and, consequently, the temperature profile of the envelope.

To be compared with the observations, the theoretical emission is
convolved with the beam pattern of the telescope. Following the
recommendations for the relevant telescope, the beam is assumed to be
a combination of gaussian curves: at 850 $\mu$m, we use HPBWs of
14.5$"$, 60$"$, and 120$"$, with amplitudes of 0.976, 0.022, and 0.002
respectively; at 450 $\mu$m, the HPBWs are 8$"$, 30$"$, and 120$"$
with amplitude ratios of 0.934, 0.06, and 0.006, respectively
\citep{San01}; at 350 $\mu$m, we use HPBWs of 12$"$ and 22$"$, with
amplitude ratios of 0.7, 0.3, respectively \citep{Hun96}.

We assumed that the envelope density follows a power law:
\begin{equation} \label{density_power_law}
  n(r)=n_0 \times \left( \frac{r_0}{r} \right)^\alpha
\end{equation}
where the power law index, $\alpha$, is a free parameter of the model,
as well as the density $n_0$, the density at $r_0$. Besides, the
envelope starts at a radius R$_{in}$ and extends up to R$_{out}$. Both
R$_{in}$ and R$_{out}$ are free parameters of the model. The last
input to DUSTY is the temperature of the central source, T$_*$, here
assumed to be 5000 K. We verified that the choice of this last
parameter does not influence the results.  Finally, the opacity of the
dust as function of the wavelength is another parameter of
DUSTY. Following numerous previous studies
\citep{Van99,Eva01,Shi02,You03}, we adopted the dust opacity
calculated by \citet{Oss94}, specifically their OH5 dust model, which
refers to grains coated by ice.

In summary, the output of DUSTY depends on $\alpha$, $n_0$, R$_{in}$
and R$_{out}$. In practice, the DUSTY input parameters are the power
law index, $\alpha$, the optical thickness at 100 $\mu$m,
$\tau_{100}$, the ratio between the inner and outer radius, Y
(=R$_{out}$/R$_{in}$) and the temperature at the inner radius
T$_{in}$. The optical thickness is, in turn, proportional to the dust
column density which depends on $n_0$ and the physical thickness of
the envelope.  Note that, since the beam of the available maps are
relatively large ($\geq 7.5"$ which corresponds to a linear length of
$\geq 3300$ AU), the inner region of the envelope is relatively
unconstrained by the available observational data. In practice, we
obtain a lower limit to T$_{in}$ of 300 K: any larger value would give
similar results.  Finally, as explained in \citet{Ive97}, DUSTY gives
scaleless results (which makes it very powerful because the same grid
of models can be applied to different sources). This means that to
compare the DUSTY output with actual observations, it is necessary to
scale the output by the source bolometric luminosity L$_{bol}$ and the
distance. Note that the bolometric luminosity is in fact estimated by
integrating the emission over the full spectrum. By definition, this
can only be done when the entire SED is known. This is exactly one of
the outputs of the modeling. So we re-evaluated the luminosity of FIR4
iteratively from the best fit model, by minimizing the
$\chi^2$\footnote{Note that, in the case of OMC2-FIR4, integrating the
  model SED gives the same results than integrating under the observed
  SED.}. We anticipate here that the new value is 1000 rather than 400
L$_\odot$, where we assumed the most recent estimation of the
distance, namely
($437\pm19$) pc \citep{Hir07}.\\

We run a grid of models to cover the parameter space as reported in
Table \ref{DUSTY_input}. The same grid of models were run for four
values of the illuminating FUV field : G$_0$ = 1, 10, 100 and 1000.
In all cases, we used the Inter-Stellar Radiation Field (ISRF)
constructed by \citet{Eva01} : combination of the radiation field
introduced by \citet{Bla94} with that of \citet{Dra78}.  Note that,
since DUSTY makes the assumption of isotropic scattering, the computed
MIR emission is largely overestimated in presence of strong
external fields (Elitzur, private communication). To solve this
problem, we followed the suggestion by \citet{You05} to neglect the
scattering, artificially putting it to zero.
\begin{table}[h] \centering
\begin{tabular}{|ll|} \hline \hline 
Parameter   & Range    \\ \hline
$\alpha$    & 0.2-3.9  \\ 
Y           & 100-2200 \\ 
$\tau_{100}$ & 0.1-4.6  \\
T$_{in}$     & 300 K   \\
T$_*$       & 5000 K   \\ \hline 
\end{tabular}
\caption{Range of the input parameters to DUSTY covered in the present
  study. The range of the $\alpha$, Y and $\tau_{100}$ parameters
  is covered by increasing by 20\% their respective value at each
  step of the grid. Note that T$_{in}$ and T$_{_*}$ are kept fixed
  as they do not influence the results (see text). \label{DUSTY_input}}
\end{table}
The best fit model has been found minimizing the $\chi^2$ with an
iterated two-steps procedure.  First, we use the observed brightness
profiles at 350, 450 and 850 $\mu m$ to constrain $\alpha$ and Y,
assuming a value for $\tau_{100}$.  Second, we constrain the 
optical thickness $\tau_{100}$ by comparing the computed and
observed SED, assuming the $\alpha$ and Y of the previous step. The
new $\tau_{100}$ is used for a new iteration and so on. In practice,
the iteration converges in two steps. This is because the normalized
brightness profiles very weakly depend on $\tau_{100}$, while they
very much depend on the sizes of the envelope and on the slope of the
density profile (see also \citet{Jor02} and \citet{Sch02}). On the
contrary, the optical thickness depends mostly on the
absolute column density of the envelope, constrained by the SED.

\subsection{Results}\label{sec:dust-results} 
 We run four grids of models, as discussed separately below: a)
  with a standard illumination FUV field (G$_o$=1) and b) with a
  10,100,1000 times enhanced field (G$_o$=10,100,1000) (see
  Introduction). In paragraph c), we also discuss why larger G$_o$
  were not considered, and in paragraph d) we summarize the results.\\

\begin{figure*} \centering
\rotatebox{90}{\includegraphics[width=12cm]{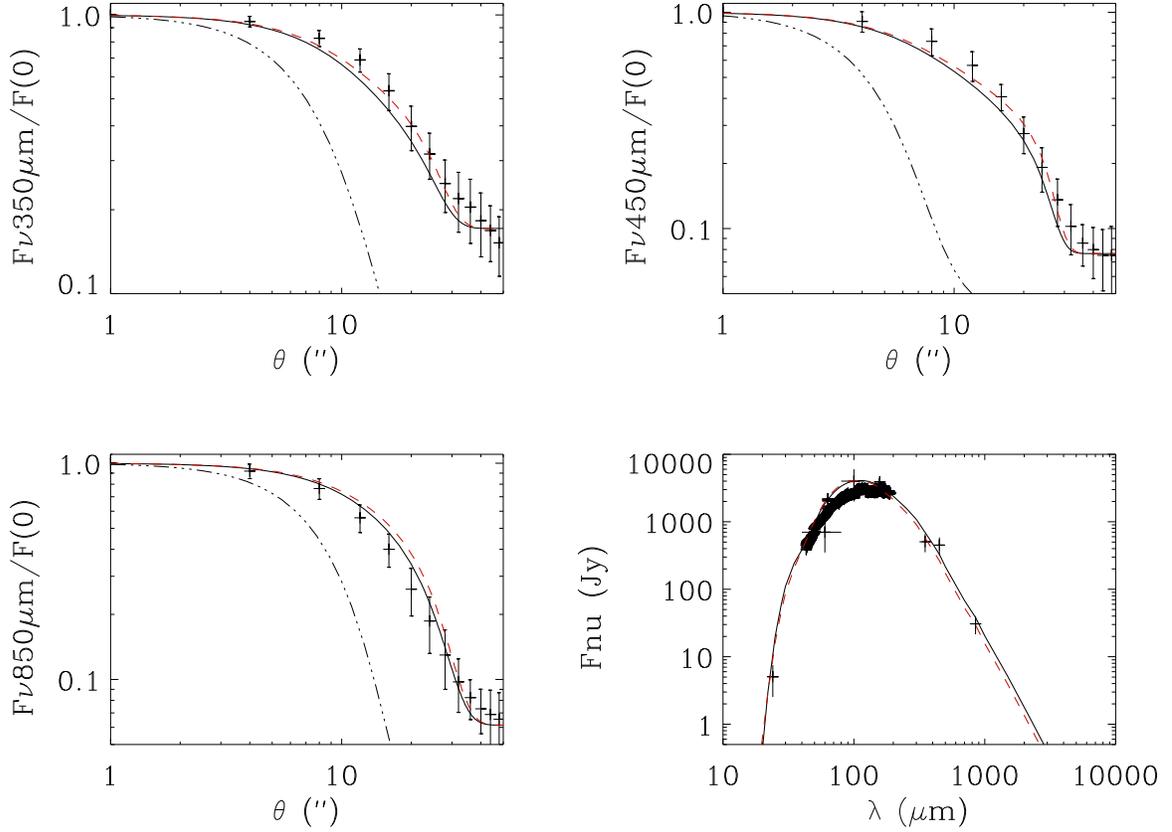}}
\caption{Observed brightness profiles at 350 (upper left panel), 450
  (upper right panel), 850 $\mu$m (lower left panel) and SED (lower
  right panel).  The curves report the best fit obtained in the two
  cases G$_o$=1 (solid line) and 1000 (dashed line). The dashed-dotted lines represent the beam pattern of the telescope adopted at 350 450 and 850 $\mu$m. Note that
    the SED plot reports the ISO-LWS spectrum between 45 and 200 $\mu
    m$ for completeness, although it has not been considered in the
    $\chi^2$ analysis, due to the relative larger calibration
    uncertainty compared to the IRAS data.}\label{fit}
\end{figure*}

\begin{figure*} \centering
\centerline{\rotatebox{90}{\includegraphics[width=12cm]{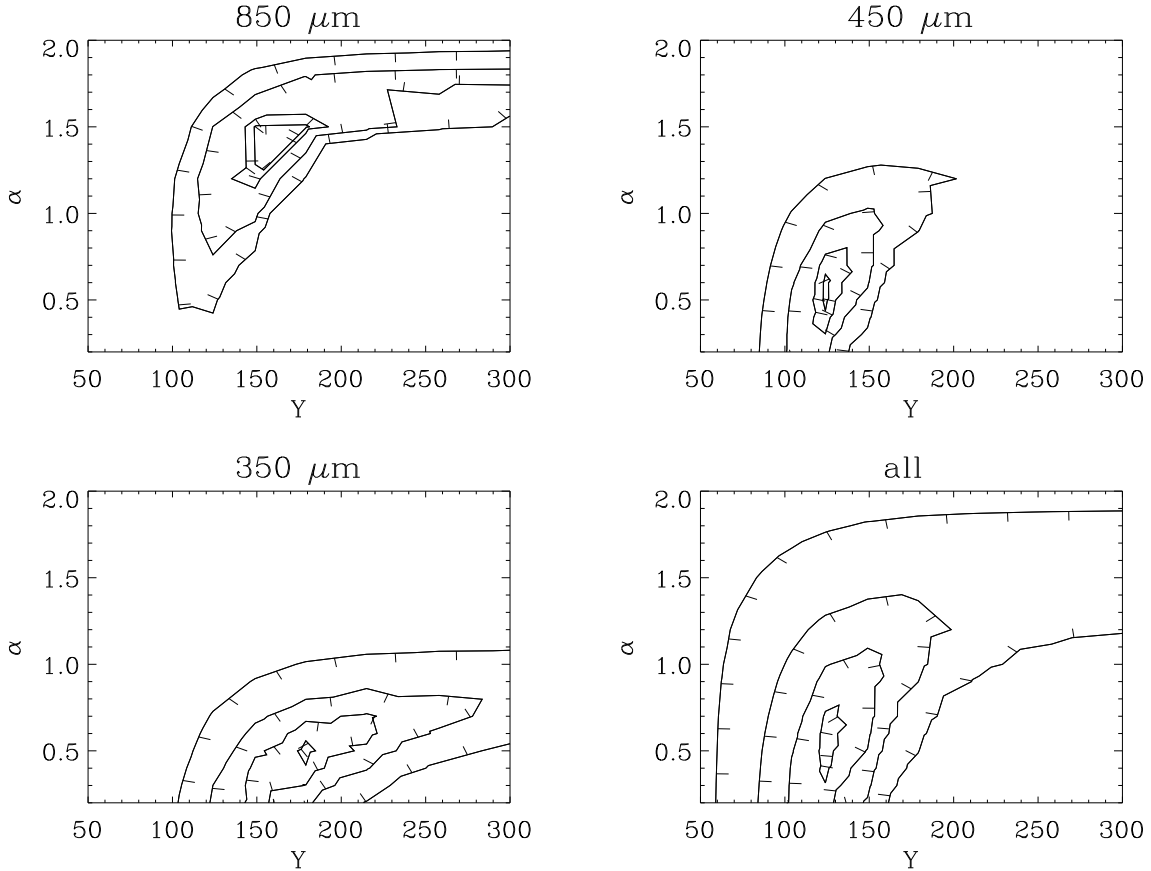}}}
\caption{$\chi ^{2}_{red}$ contour plots (Y,$\alpha$) for the models
  with G$_{o}$=1. In these computations, $\tau_{100}$ is equal to
  0.6. The contours show the loci of the $\chi ^{2}_{red}$ values
    equal to 1.1, 1.5, 2.5 and 5 times the minimum $\chi ^{2}_{red}$.
 The upper left panel is obtained by comparing the model predictions
  with the 850 $\mu$m brightness profile; the upper right panel refers
  to the 450 $\mu$m profile; the lower left panel refers to the 350
  $\mu$m profile; the lower right panel makes use of the three
  profiles. \label{fig:chi2-contour}}
\end{figure*}

\begin{figure} \centering
\centerline{\rotatebox{0}{\includegraphics[width=10cm]{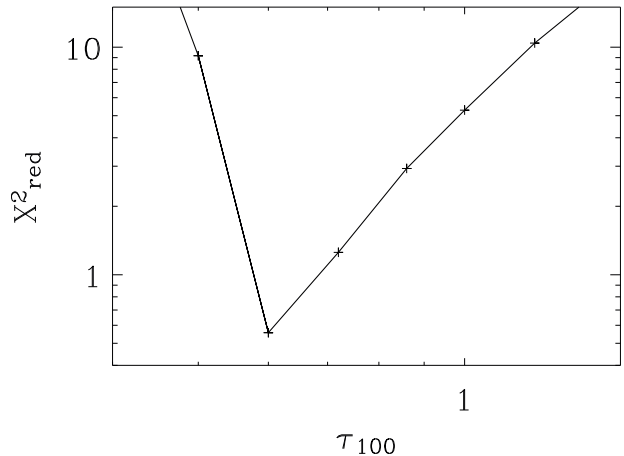}}}
\caption{$\chi^2_{red}$ versus $\tau_{100}$. In these computations, Y
  is equal to 120 and $\alpha$ is equal to 0.6. \label{X2SED}}
\end{figure}

\noindent
{\it a) G$_o$=1}\\
Table \ref{best_fit_no_ISRF} presents the set of parameters $\alpha$,
Y and $\tau_{100}$, which better reproduce the observations 
assuming G$_o$=1. Figure \ref{fit} shows the relevant derived
brightness profiles and SED against the observed ones. Figure
\ref{fig:chi2-contour} shows the $\chi^2$ contours plots obtained by
considering separately the brightness profiles at 350, 450 and 850
$\mu m$, and by combining the three profiles together. Figure
\ref{X2SED} shows the $\chi^2$ dependence on the $\tau_{100}$
parameter.
\begin{table}[h] \centering
\begin{tabular}{l|ccc|cc} \hline \hline 
Observation  & $\alpha$ & Y & $\tau_{100}$ & $\chi _{red}^{2}$ & $\nu$\\ \hline
850 $\mu$m profile  & 1.4 & 160 & -    & 0.72 & 10 \\
450 $\mu$m profile  & 0.6 & 120 & -    & 0.63 & 10 \\
350 $\mu$m profile  & 0.5 & 170 & -    & 0.47 & 10 \\ 
All profiles        & 0.6 & 120 & -    & 1.24 & 36 \\ 
SED                 & -   & -   & 0.6  & 0.55 &  3 \\ \hline
\end{tabular}
\caption{Best fit parameters for the case G$_o$=1. Note that $\chi
  _{red}^{2} = \chi ^{2}/\nu$ where $\nu$ is the number of degrees of
  freedom. The first line reports the best fit obtained using only the
  850 $\mu$m brightness profile; second line, using the 450 $\mu$m
  brightness profile; third line, using the 350 $\mu$m brightness
  profile; fourth line gives the best fit using the three profiles;
  the last line gives the best fit using the
  SED.\label{best_fit_no_ISRF}}
\end{table}

The three $\chi _{350}^{2}$, $\chi _{450}^{2}$ and $\chi _{850}^{2}$
contour plots point to a value of Y around 100-200. Conversely, the
$\chi _{350}^{2}$ and $\chi _{450}^{2}$ contour plots constraint
$\alpha$ to a lower value than 1, around 0.5-0.6, whereas the $\chi
_{850}^{2}$ would rather indicate a larger value for $\alpha$,
although the value 0.6 is still acceptable. Note that the solution
found by \citet{Jor06} relies on the 850 $\mu$m profile only, and,
therefore, gives a large $\alpha$ value, consistent with our
  $\chi _{850}^{2}$ plot. The $\chi _{SED}^{2}$ plot
(Fig.\ref{X2SED}) points to a value of $\tau_{100}$ of 0.6. In
  minimizing the $\chi _{SED}^{2}$, we varied the source luminosity
  from 400 to 1500 L$_\odot$. The best fit is
  obtained for a source luminosity equal to 1000 L$_\odot$.\\

\begin{figure*} \centering
\rotatebox{90}{\includegraphics[width=6cm]{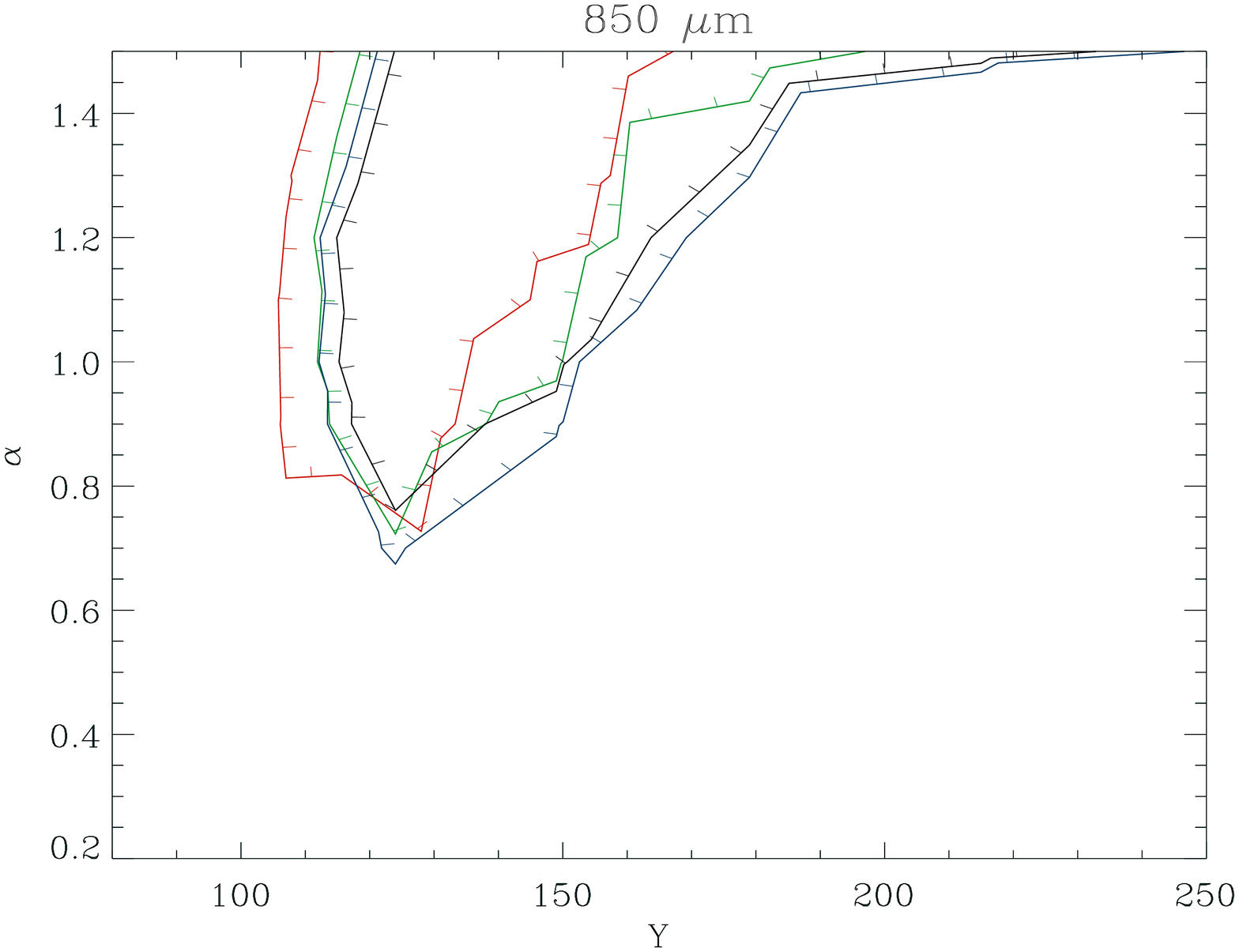}}
\rotatebox{90}{\includegraphics[width=6cm]{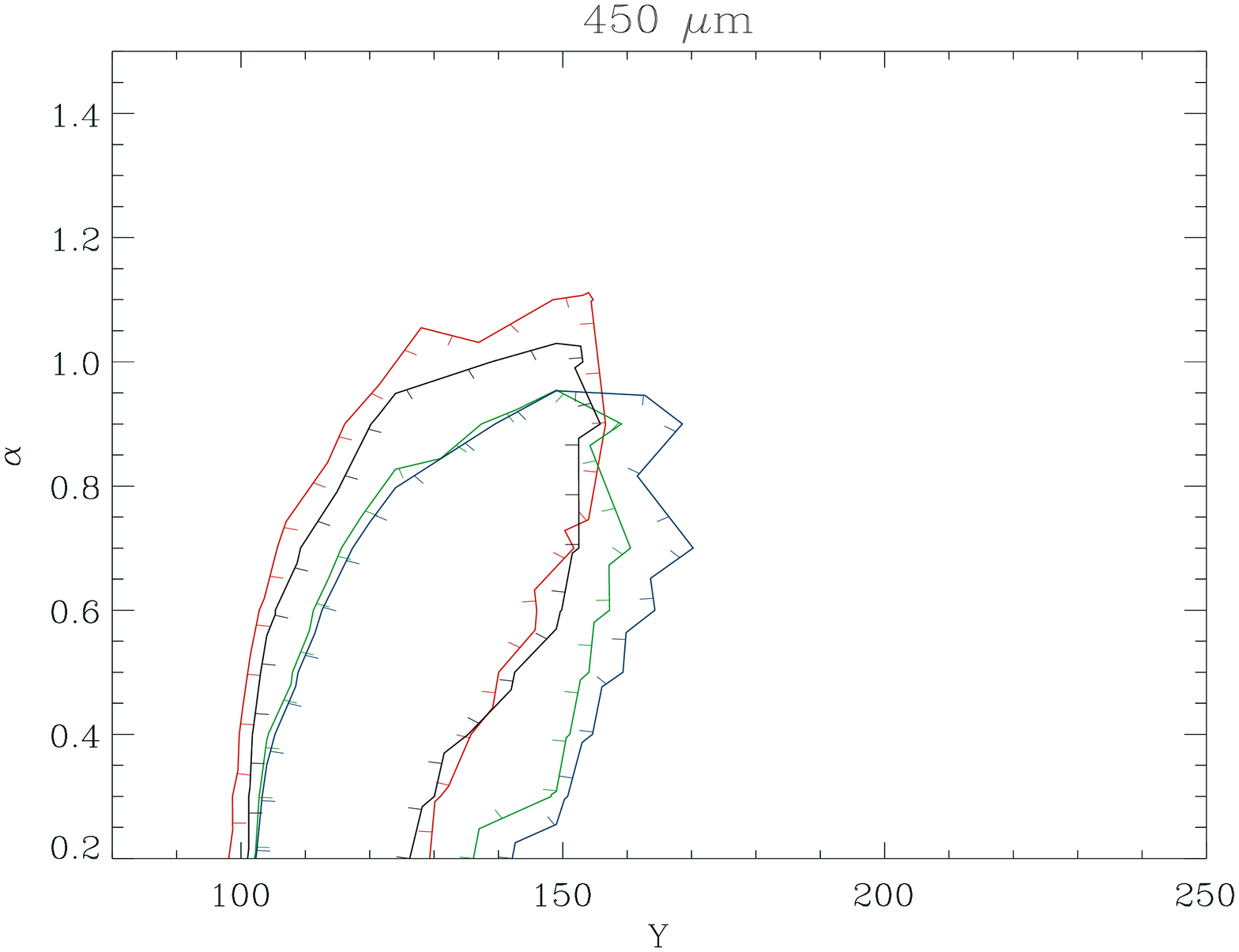}}
\rotatebox{90}{\includegraphics[width=6cm]{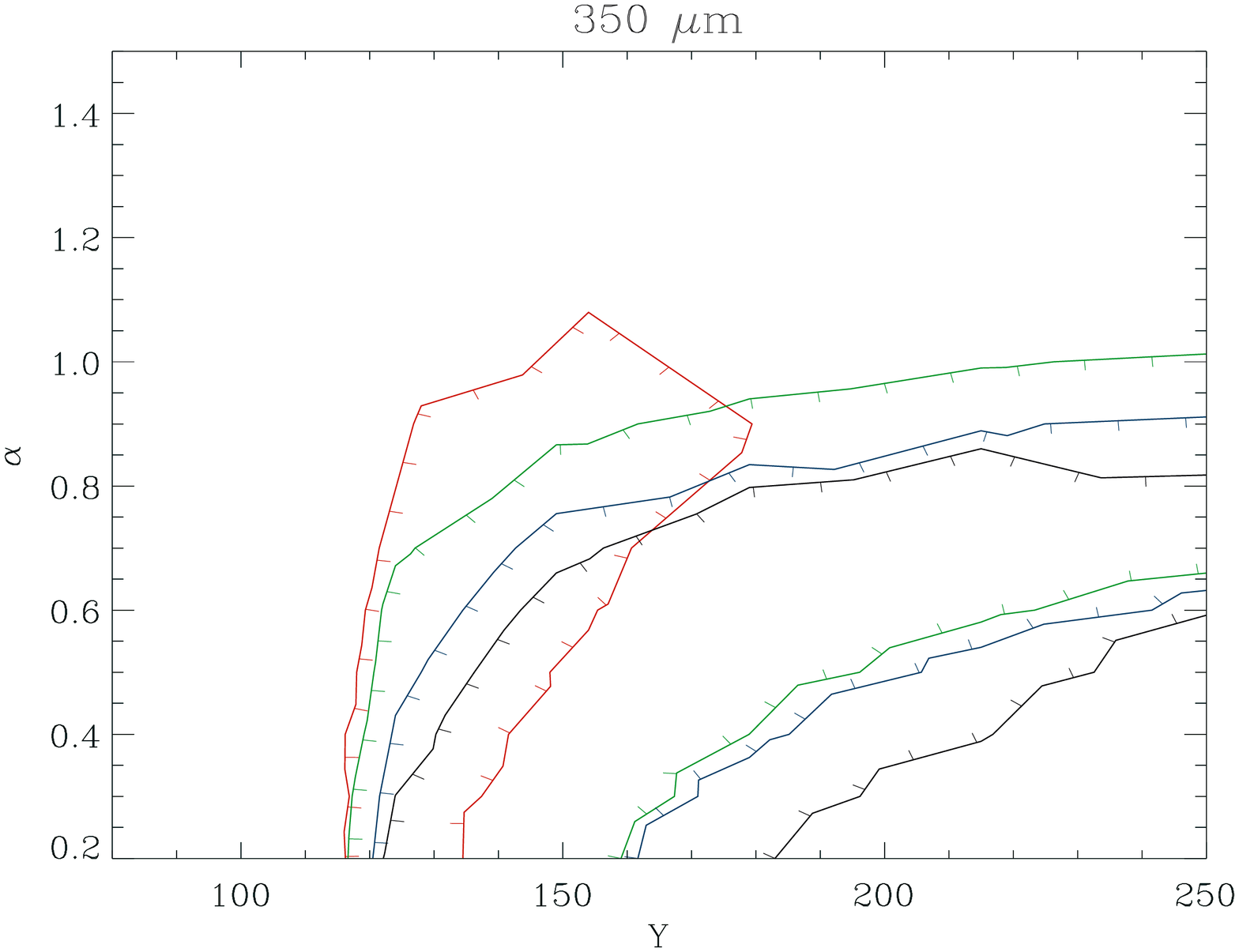}}
\rotatebox{90}{\includegraphics[width=6cm]{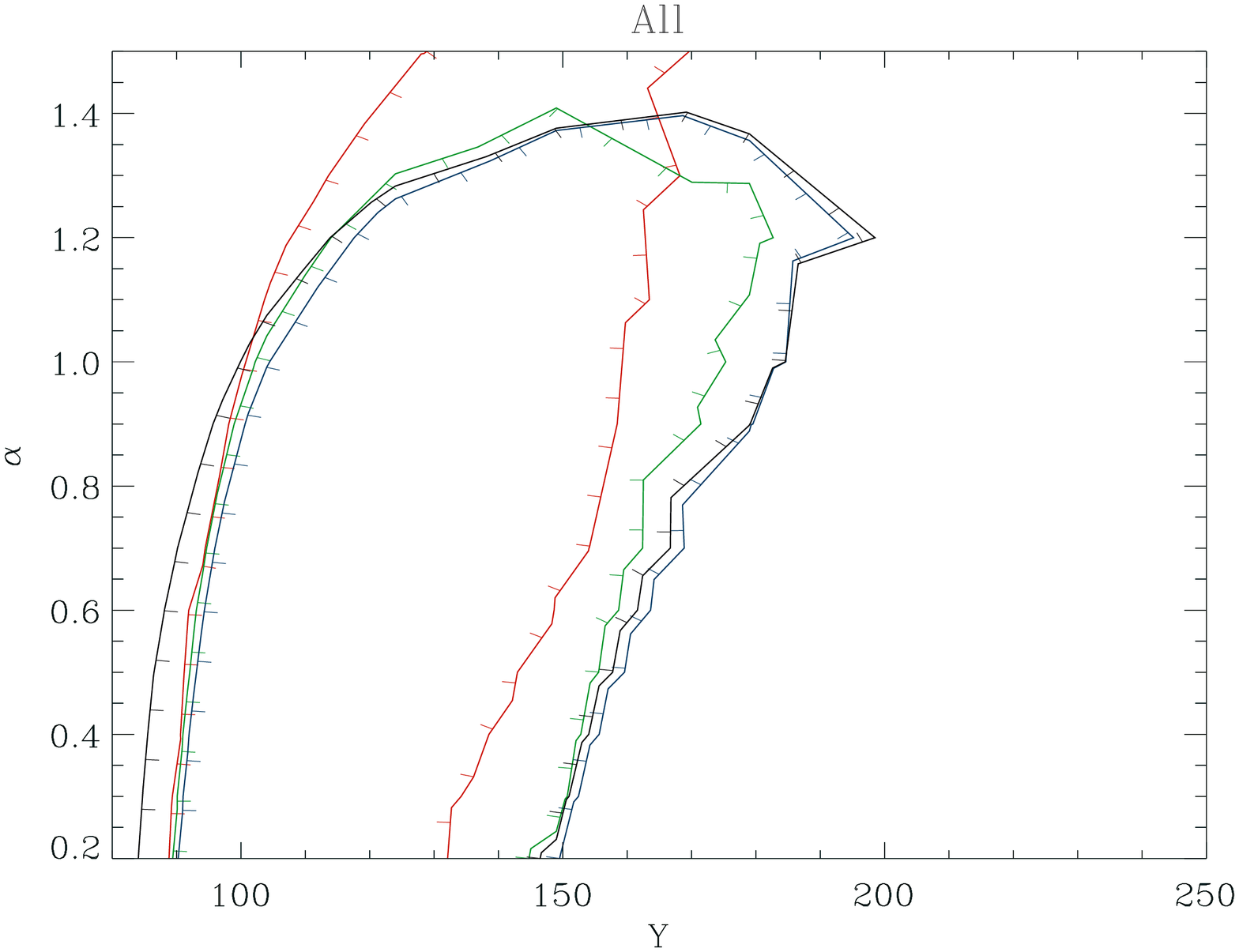}}
\caption{$\chi ^{2}_{red}$ contour plots (Y,$\alpha$) for the models
  with G$_{o}$=1 (black lines), 10 (blue lines), 100 (green lines) and
  1000 (red lines). In these computations, $\tau_{100}$ is equal to
  0.6. The contours show the loci of the $\chi ^{2}_{red}$ values
    equal to 2.5 times the minimum $\chi ^{2}_{red}$. From top to
  bottom: $\chi^2$ contours of the 850 $\mu$m, 450 $\mu$m, 350 $\mu$m
  and the three together.\label{X2all_ISRF}}
\end{figure*}

\noindent
{\it b) G$_o$=10,100,1000}\\
The best fit values of $\alpha$ and Y for cases of an enhanced
illumination UV field are presented in Figure \ref{X2all_ISRF}.
The first thing to notice is that the $\chi _{all}^{2}$ does not
change appreciably for G$_o$ equal to 1,10,100 or 1000: the minimum
$\chi _{all_{red}}^{2}$ value is 1.24, 1.23, 1.19 and 1.20 for
G$_o$=1, 10, 100 and 1000 respectively. In other words, the available
continuum observations, both the profiles and the SED, cannot
distinguish which of the four models is better. Furthermore, 
Figure \ref{X2all_ISRF} shows that the $\chi _{all}^{2}$ contour
plots point to the same Y and $\alpha$ values. Similarly, the
$\tau_{100}$ value is 0.6 for the four cases G$_o$=1, 10, 100 and
1000.  The situation is illustrated in Fig. \ref{fit}, where the best
fit predictions are compared to the observations for the two cases
G$_o$=1 and 1000. Both models reproduce fairly well the observations,
as implicit in the best-$\chi^2$ similar values. Note, however, that
the G$_o$=1000 case predicts slightly larger fluxes, due to the
enhanced temperature at the border of the envelope.\\

\noindent
{\it c) larger G$_o$ }\\
 We did not explore in detail the case of larger G$_0$ for three
  reasons. The first one is that previous line observations showed
  that the FUV field in the OMC2 region is ``only'' 500 times the
  Interstellar field. Indeed, \cite{Her97} mapped the OMC-2 cloud in
  the CII-157 $\mu$m, OI-63 and -146 $\mu$m lines with the
  spectrometer FIFI on board the Kuiper Airborne Observatory. They
  detected extended emission associated with the Photo-Dissociation
  Region (PDR) enveloping the whole OMC-2 molecular cloud. These
  authors concluded that OMC-2 is illuminated by a FUV field whose
  intensity is G$_0$ $\sim$ 500. Note that this is the FUV field
  impinging on the cloud and that the effective G$_0$ seen by the FIR4
  envelope is probably lower than this. The second reason is that
varying G$_0$ from 1 to 10$^3$ does not improve the $\chi^2$
value. The third reason is that the G$_0$=10$^4$ case suffers of
severe convergence problems, and it was not possible to derive a large
enough number of runs for a meaningful $\chi^2$
analysis.\\

\begin{figure} \centering
\rotatebox{0}{\includegraphics[width=9cm]{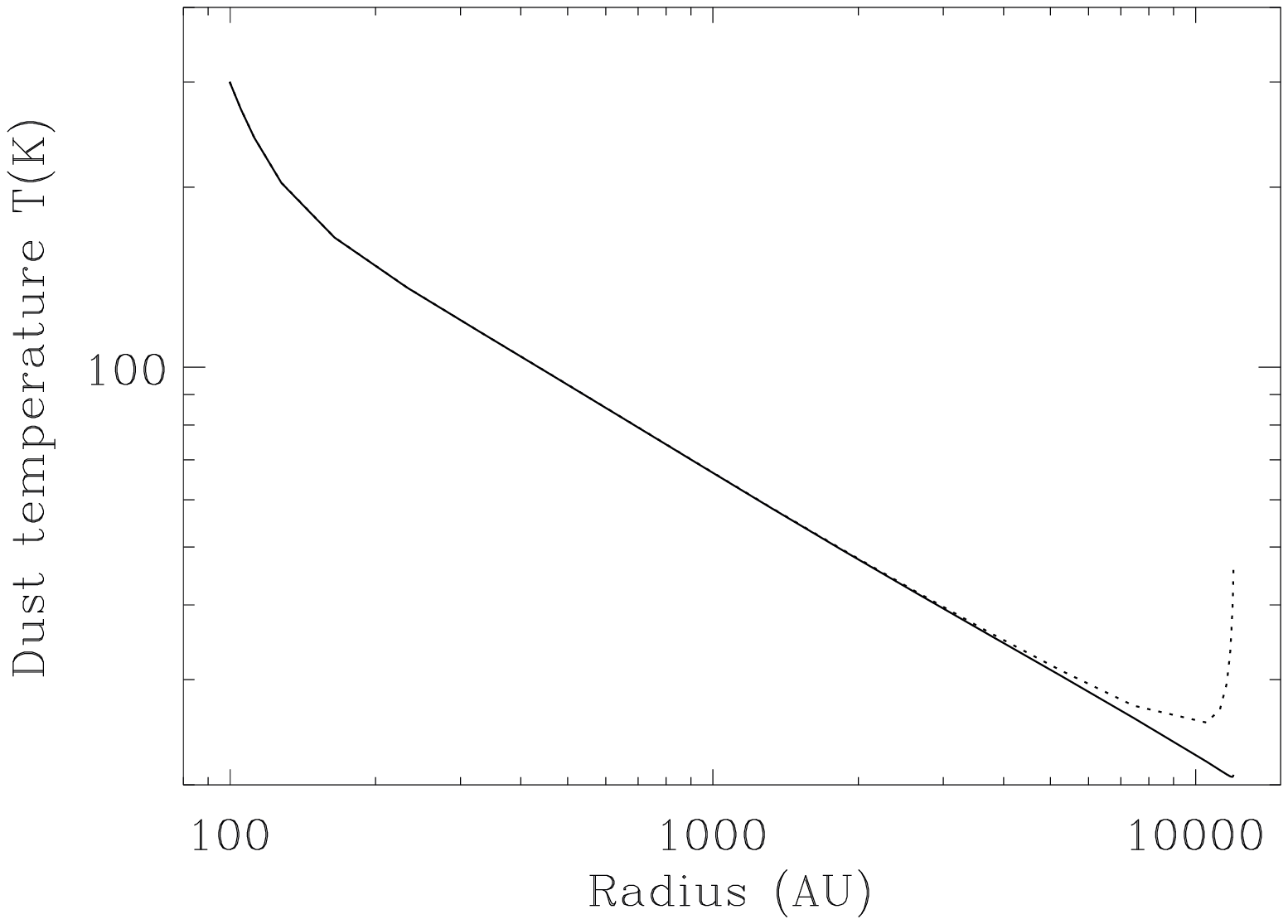}}
\rotatebox{0}{\includegraphics[width=9cm]{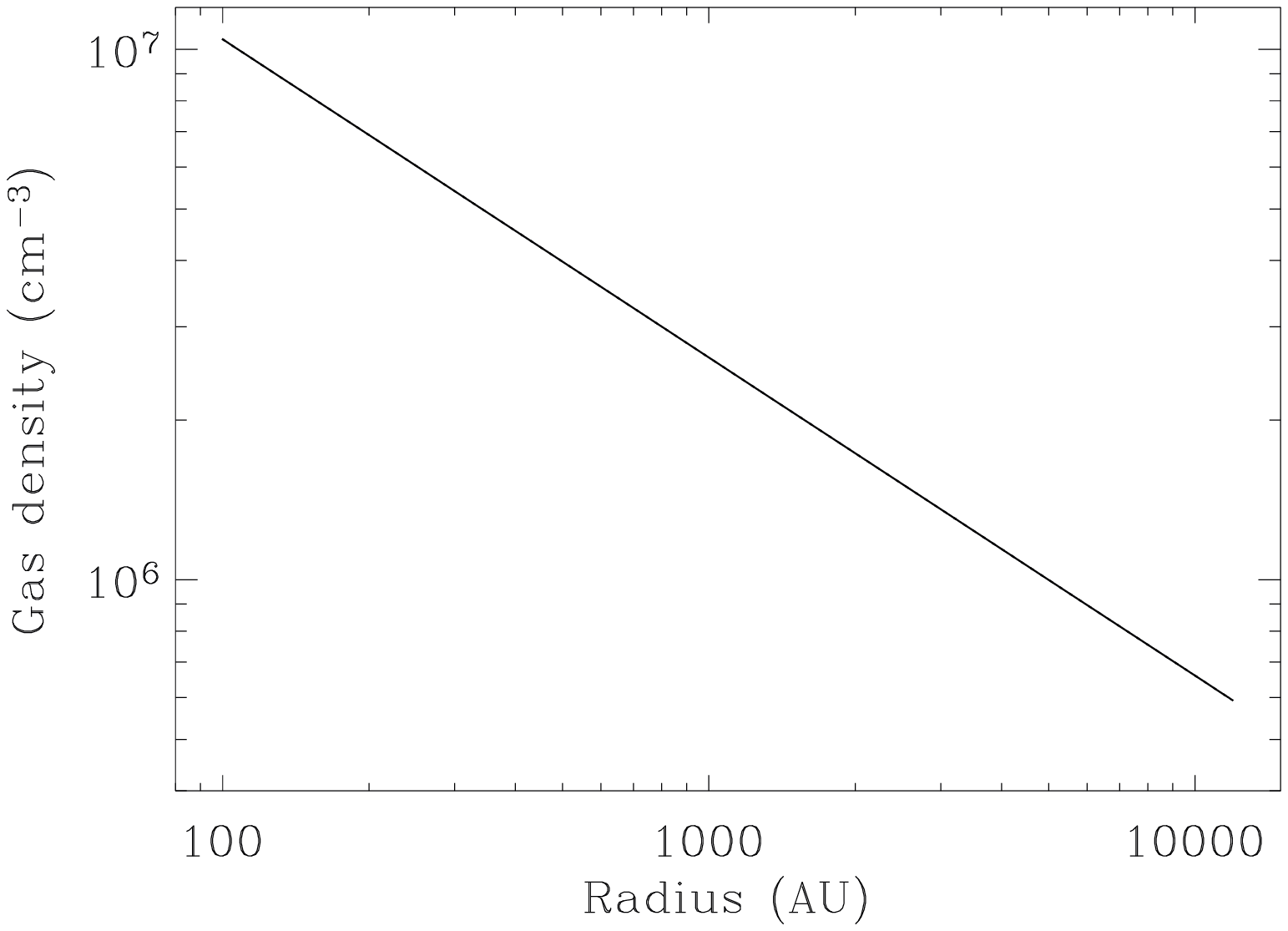}}
\caption{Dust temperature (upper panel) and H$_{2}$ density (lower panel) 
profiles from the best fit obtained in the two cases G$_0$=1 and 1000. The 
plain line and the dotted line represent the cases G$_0$=1 and 1000 
respectively.}\label{T_n.ps}
\end{figure}

\noindent
{\it d) Summary of the adopted solution}\\
 Table \ref{best_fit_phy_param} summarizes the value of the best
  fit parameters, obtained by considering all the profiles and the SED
  $\chi^2$ contour plots and assuming the G$_0$=1 case. Some relevant
  physical quantities are quoted in the same table. Fig. \ref{T_n.ps}
  shows the dust density and temperature profiles of the best fit
  models with G$_0$=1 and 1000 respectively. Note that the dust
  temperature in the skin of the envelope is larger by $\sim$ 20-30 K
  in the case G$_0$=1000 with respect to the G$_0$=1 case. This
  increase concerns a relatively small region, of a few thousand
  AU. \citet{Jor06} found a larger warm region, of about
  10$^4$ AU, because of the steeper adopted density distribution
  ($\alpha$=2): in this case, the FUV photons can penetrate deeper
  into the envelope.
\begin{table}
 $$
 \begin{array}{p{0.5\linewidth}r}
   \hline
   \multicolumn{2}{c}{\mathrm{Fixed\ input\ parameters}} \\
   \hline
   \noalign{\smallskip}
   Distance, $d$	  & 437\,\mathrm{pc}	  \\
   Stellar temperature, $T_{\star}$    & 5000\,\mathrm{K}\\
   Dust temperature at $r_{\mathrm{i}}$, $T_{\mathrm{d}}(r_{\mathrm{i}})$   & 300\,\mathrm{K}\\
   Dust opacity (OH5) at 100\,$\mu$m, $\kappa_{100}$ & 86.5\,\mathrm{cm}^2\mathrm{g}^{-1} \\
 \hline
 \noalign{\smallskip}
 \multicolumn{2}{c}{\mathrm{Best\ fit\ parameters}} \\
 \hline
 \noalign{\smallskip}
 Luminosity, $L$   & 1000\,\mathrm{L}_{\sun}\\
 Dust optical depth at 100\,$\mu$m, $\tau_{100}$     & 0.6 \\
 Density power law index, $\alpha$   & 0.6 \\
 Envelope thickness, $r_{\mathrm{out}}$/$r_{\mathrm{i}}$  & 120 \\
 \hline
 \noalign{\smallskip}
 \multicolumn{2}{c}{\mathrm{Physical\ quantities}} \\
 \hline
 \noalign{\smallskip}
 Inner envelope radius, $r_{\mathrm{in}}$ 	 & 100\,\mathrm{AU}\\
 Outer envelope radius, $r_{\mathrm{out}}$ 	 & 12000\,\mathrm{AU}\\
 Radius at T$_{dust}$ = 100 K,   $r_{\mathrm{100K}}$ & 440\,\mathrm{AU}\\
 H$_2$ density at $r_{\mathrm{100K}}$, $n_0$ 	 & 4.3\times 10^{6}\,\mathrm{cm}^{-3} \\
 Envelope mass, $M_{\mathrm{env}}$		 & 30\,\mathrm{M}_{\sun} \\
 \hline
 \noalign{\smallskip}
 \end{array}
 $$
 \caption[]{ Summary of the dust radiative transfer analysis of
     OMC2-FIR4. The first part lists the fixed input parameters, 
     the second part reports the best fit parameters, and some relevant 
     physical quantities corresponding to the
     best fit model are reported in the third part.}
 \label{best_fit_phy_param}	 
\end{table}

\section{Gas temperature profile}\label{sec:gas-temp-prof}

\subsection{Model description}
\citet{Cec96}, \citet{Dot97} and \citet{Mar02} showed that the gas is
thermally decoupled from dust in the inner regions of low and high
mass protostellar envelopes. The reason for that is the large water
abundance in the gas phase caused by the sublimation of the grain
mantles. The same phenomenon may occur in the envelopes of
intermediate mass protostars, so we explicitly computed the gas
temperature profile of the envelope surrounding FIR4. For that we
explicitly computed the equilibrium temperature by equating the gas
cooling and heating terms at each radius. Following the method
described in \citet{Cec96}, we considered heating from the gas
compression (due to the collapse), dust-gas collisions and
photo-pumping of H$_2$O and CO molecules by the IR photons emitted by
the warm dust close to the center\footnote{Cosmic rays ionization is a
  minor heating term in the protostellar envelopes.}. The cooling is
mainly due to rotational lines from H$_2$O and CO, plus the fine
structure lines from O. Therefore, the gas temperature depends on the
abundance of these three species. In practice, though, only the water
abundance is a real parameter of the model, because the CO and O lines
are optically thick in the range of the CO and O abundances typical of
protostellar envelopes. For this reason, we computed various cases for
the water abundance, as it is generally poorly constrained in
protostellar envelopes, and totally unconstrained in FIR4 (see \S
\ref{sec:predicted-water-line}).  We adopted a step function for the
water abundance profile to simulate the jump caused by the ices
sublimation. The jump is assumed to occur at 100 K. We considered the
H$_2$O abundance (with respect to H$_2$) X(H$_2$O)$_{out}$ in the
outer envelope, where T$\leq$ 100 K, equal to 10$^{-7}$, 10$^{-8}$ and
10$^{-9}$.  We also considered three cases for the abundance in the
inner region X(H$_2$O)$_{in}$, 10$^{-4}$, 10$^{-5}$ and
10$^{-6}$. Finally, we studied the case with G$_0$ = 1000. The run
parameters are summarized in Table \ref{tab:abu-cool}.
\begin{table}[tbh]
  \centering
  \begin{tabular}{|cccccc|}
    \hline
    Model & X(CO)            & X(O)             & X(H$_2$O)$_{out}$& X(H$_2$O)$_{in}$ & G$_0$\\ \hline
    1$^a$ & $1\times10^{-4}$ & $5\times10^{-4}$ & $1\times10^{-8}$ & $1\times10^{-5}$ & 1\\
    2     & $1\times10^{-4}$ & $5\times10^{-4}$ & $1\times10^{-8}$ & $1\times10^{-6}$ & 1\\
    3     & $1\times10^{-4}$ & $5\times10^{-4}$ & $1\times10^{-8}$ & $1\times10^{-4}$ & 1\\
    4     & $1\times10^{-4}$ & $5\times10^{-4}$ & $1\times10^{-9}$ & $1\times10^{-5}$ & 1\\
    5     & $1\times10^{-4}$ & $5\times10^{-4}$ & $1\times10^{-7}$ & $1\times10^{-5}$ & 1\\
    6     & $1\times10^{-4}$ & $5\times10^{-4}$ & $1\times10^{-7}$ & $1\times10^{-4}$ & 1\\
    7     & $1\times10^{-4}$ & $5\times10^{-4}$ & $1\times10^{-8}$ & $1\times10^{-5}$ & 1000\\ \hline
  \end{tabular}
  \caption{The different run models (column 1). Column 2 to 5 report 
    the adopted abundances of the main gas coolants: X(CO), X(O) and the 
    H$_2$O abundance in the T$\geq$100 K region X(H$_2$O)$_{in}$ and outer 
    region X(H$_2$O)$_{out}$. Column 5 reports the FUV illuminating 
    field G$_0$. Note: $^a$ Model 1 is the reference for the studies of
    the water line spectrum presented in \S \ref{sec:predicted-water-line}. 
    \label{tab:abu-cool}}
\end{table}

To compute the cooling from the lines we used the code described in
\citet{Cec96,Cec03} and \citet{Par05}. The same code has been used in
several past studies, whose results have been substantially confirmed
by other groups (e.g. the analysis on IRAS16293-2422 by
\citep{Sch02}). Briefly, the code is based on the escape probability
formalism in presence of warm dust (see \citet{Tak83}), where the
escape probability $\beta$ is computed at each point by integrating
the line and dust absorption over the solid angle $\Omega$ as follows:
\begin{equation} 
  \beta = \frac{k_\mathrm{d}}{k_\mathrm{L} + k_\mathrm{d}} + 
  \frac{k_\mathrm{L}}{(k_\mathrm{L} + k_\mathrm{d})^2} \int d\mu 
  \frac{1-\exp \left[ - \left( k_\mathrm{L} + k_\mathrm{d} \right) 
  \Delta L_\mathrm{th} \right]} {\Delta L_\mathrm{th}} 
\end{equation} 
\noindent  
where $k_\mathrm{L}$ and $k_\mathrm{d}$ are the line and dust
absorption coefficients respectively, and $\Delta L_\mathrm{th}$ is
the line trapping region, given by the following expressions:
\begin{equation} 
  \Delta L_\mathrm{th} = 2 \Delta v_\mathrm{th}  
  \left( \frac{v}{r} \left| 1-\frac{3}{2} \mu^2 \right| \right)^{-1}  
\end{equation} 
\noindent
in the infalling region of the envelope (where $\mathrm{arcos} \left(
\mu \right)$ is the angle with the radial outward direction) and
\begin{equation} 
  \Delta L_\mathrm{th} = r \left( 1 - \frac{r}{R_\mathrm{env}} \right) 
\end{equation} 
\noindent 
in the static region (where $R_\mathrm{env}$ is the envelope radius).
 In the present calculations, we assumed that the entire envelope is
collapsing in free-fall towards a central object of 2 M$_{\sun}$.  In
practice, the photons emitted by the dust can be absorbed by the gas
and can pump the levels of the water molecules. This, indeed, is an
important factor in the population of the water levels, and, for the
highest energy levels, even the dominant one (\S
\ref{sec:predicted-water-line}). In addition, H$_2$O and CO
  molecules can be pumped by absorption of the NIR photons emitted by
  the innermost warm dust. Since the densities and temperatures of the
  regions of the envelope targeted by this study are not enough to
  populate the levels at the vibrational states, the effect of the NIR
  photons is an extra heating of the gas, as described in the
  Ceccarelli et al. (1996) article. Note that the code takes into
  account the dust with temperatures up to 1500 K, by following the
  algorithm described in Ceccarelli et al. (1996).

For the collisional coefficients of water with hydrogen molecules, we
used the data by \citet{Fau07} available for the temperature range
20-2000K. This data set includes quasi-classical results for the
highest rates (those larger than 10$^{-12}$ cm$^3$s$^{-1}$) and quantum
scaled H$_2$O-He results for the lowest rates. Recent quantum
calculations on ortho-H$_2$O by Dubernet and co-workers have shown
that the quasi-classical rates can be in error by as much as a factor
of 100 but that, in general, they are accurate to within a factor of
1-3 \citep{Dub09}. It should be noted that
the rates of \citet{Fau07} are currently the only complete and
consistent set of data for both ortho- and para-H$_2$O colliding with
H$_2$. We also note that these rates have been recently extrapolated
in order to cover energy levels and temperatures up to 5000K
\citep{Fau08}. Since the ortho to para conversion process of H$_2$ is
chemical rather than radiative, the ortho-to-para ratio H$_2$ OPR is
highly uncertain in the interstellar medium. Here we assume that in
warmer gas it is in Local Thermal Equilibrium and, therefore, follows
the Boltzmann distribution:
\begin{equation}
OPR =  \frac{(2I_o+1) \Sigma (2J+1) \exp(-\frac{E_o(J)}{kT})}
            {(2I_p+1) \Sigma (2J+1) \exp(-\frac{E_p(J)}{kT})}
\end{equation}
where $I_o$ and $I_p$ are the total nuclear spin, corresponding to
whether the hydrogen nuclear spins are parallel ($I_o = 1$,
$\uparrow\uparrow$) or anti-parallel ($I_p = 0$,
$\uparrow\downarrow$). The sum in the numerator and denominator
extends over all ortho and para levels J, respectively. Similarly to
H$_2$, water comes in the ortho and para forms. In this case, since
the water is the dominant gas coolant only in the regions where the
dust temperature exceeds 100 K, we assumed OPR equal to 3, strictly
valid for gas temperatures larger than 60 K.  Since the water lines
are optically thick, the cooling depends on the velocity field,
assumed to be that of an envelope collapsing in free-fall towards a
central object of 2 M$_{\sun}$ (see above). We checked the influence
of our results against this assumption, running a case with a constant
velocity field of 0.5 km/s. The difference in the gas temperature
between the two cases never exceeds 10 K.

\subsection{Results}\label{sec:results-gastemp}
\begin{figure}
\rotatebox{0}{\includegraphics[width=9cm]{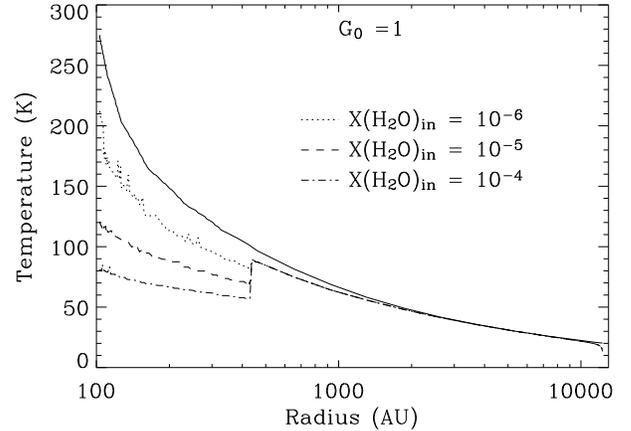}}
\caption{The gas temperature profile of the collapsing envelope of
  OMC2-FIR4. The different curves refer to different values of the
  inner envelope water abundance X(H$_2$O)$_{in}$: $1\times10^{-6}$
  (dotted), $1\times10^{-5}$ (dashed) and $1\times10^{-4}$
  (dotted-dashed) respectively. In these computations,
  X(H$_2$O)$_{out}$ is $1\times10^{-8}$ and G$_0$=1. The solid line
  refers to the dust temperature profile.\label{tg.ps}}
\end{figure}

Figure \ref{tg.ps} shows the computed gas temperature profile obtained
with different values of X(H$_2$O)$_{in}$ in the case G$_0$ = 1.
\begin{figure}
\rotatebox{0}{\includegraphics[width=9cm]{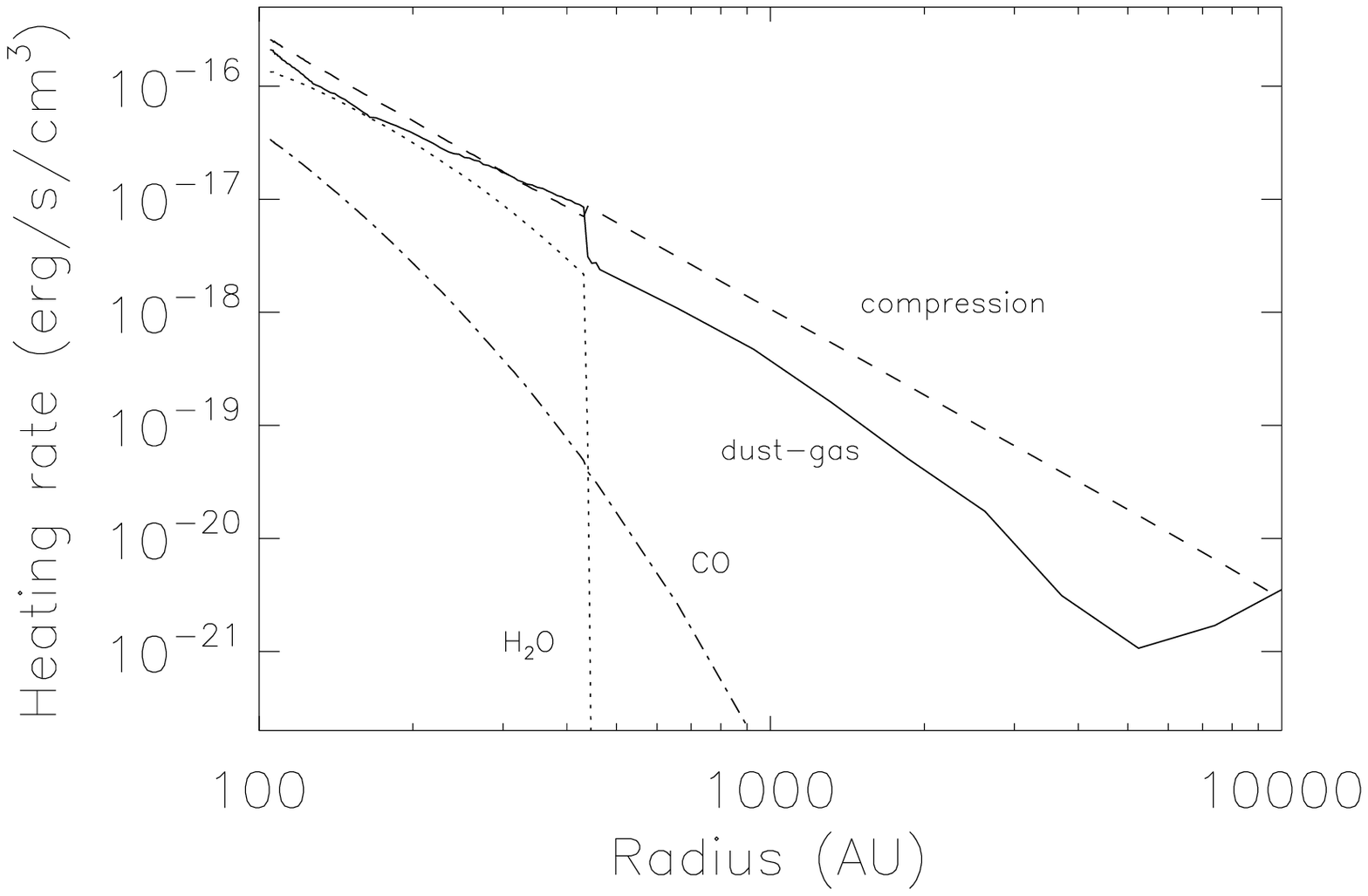}}
\rotatebox{0}{\includegraphics[width=9cm]{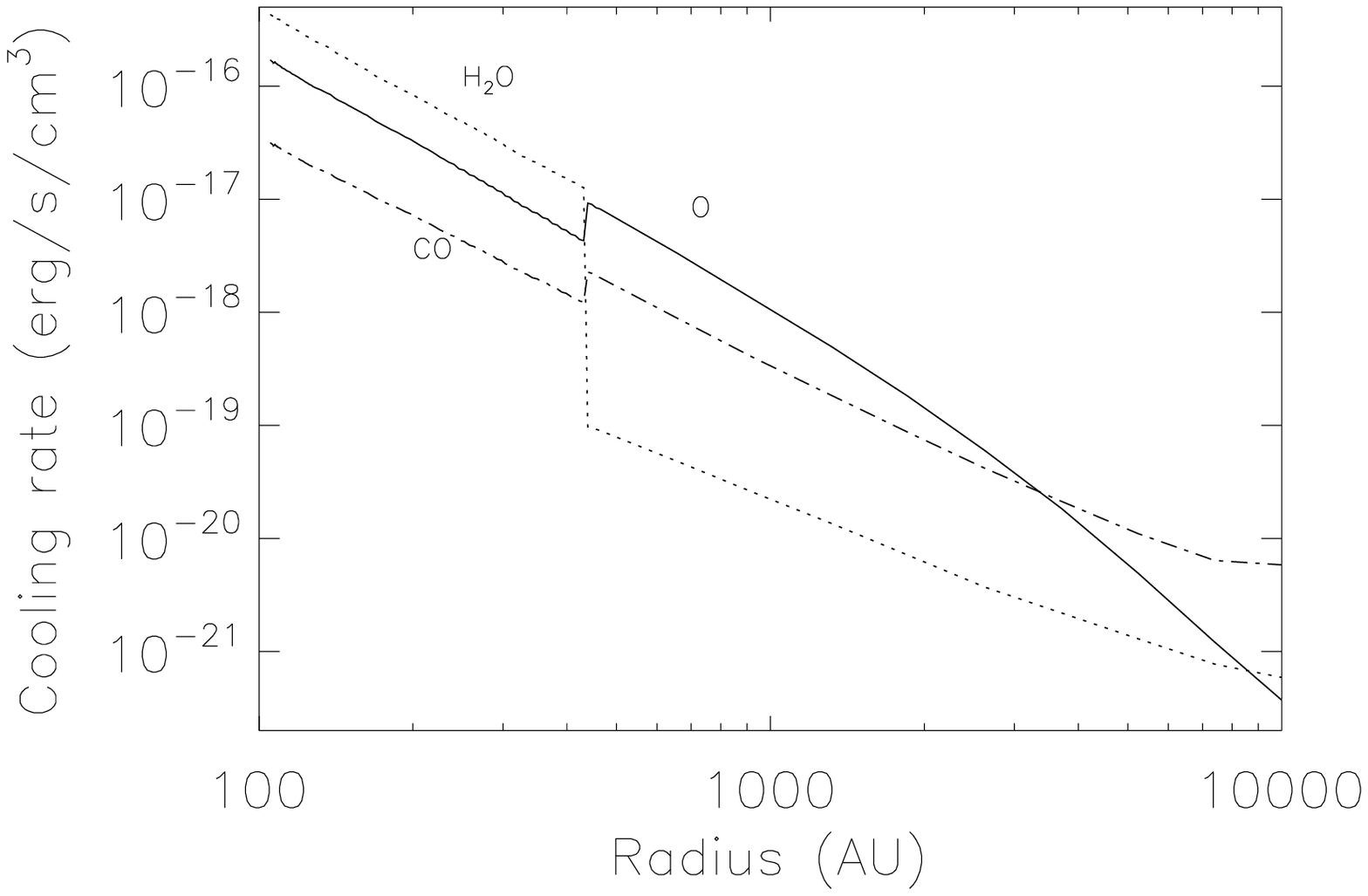}}
\caption{Heating (top panel) and cooling (bottom panel) rates as
  function of the radius, computed assuming that the inner H$_2$O
  abundance is equal to $1\times10^{-5}$ while the outer abundance is
  $1\times10^{-8}$. \label{thermal.ps}}
\end{figure}
 Figure \ref{thermal.ps} shows the different contributions to the
  heating and cooling rates. Similarly to what had been found in low
mass protostars \citep{Cec96,Cec00,Mar02}, the gas temperature tracks
the dust temperature in the outer envelope while gas and dust are
decoupled in the inner part of envelope, where the icy grain mantles
sublimate. The heating is dominated by compression of the
  collapsing gas across the entire envelope, even though the dust-gas
  collisions becomes comparable to the compression heating in the
  inner envelope. Although important in the very inner regions, the
  H$_2$O photo-pumping never dominates the heating contrarely to what
  happens in the studied low mass protostars. The cooling, on the
  other hand, is dominated by H$_2$O line emission in the inner
  envelope, by the OI line emission in the intermediate region and by
  the CO in the outermost regions of the envelope.
Note that the increased water abundance causes an increased cooling
of the gas, which brings the equilibrium gas temperature to lower
values than the dust temperature. This phenomenon, already predicted
in low mass protostars, is much more marked in the FIR4 case, leading
to more than 100\% of difference (with respect to the gas temperature)
in the dust and gas temperatures for the case of the highest water
abundance ($1\times10^{-4}$). For example, at 100 AU the dust
temperature is 300 K, whereas the gas temperature varies from 200 to
80 K depending on the assumed X(H$_2$O)$_{in}$, $1\times10^{-6}$ and
$1\times10^{-4}$ respectively. The phenomenon is more marked in FIR4
than in the studied low mass protostars because of the relatively
lower density of the region where the icy grain mantles sublimate in
FIR 4 than in the low mass protostars, or, in other words, because the
FIR4 envelope is warmer.  Note that we obtain similar results also for
larger illuminating FUV fields.

 We emphasize that this result is a consequence of the derived
  shallow dependence of the density distribution, which is constrained
  from the fit of the maps. The dependence is strictly valid only at
  scales larger than the smaller telescope beam, namely 8$"$
  (equivalent to a radius of about 1700 AU) and the SED fit only gives
  the total column density, which, coupled with the density dependence
  on the radius (constrained by the maps), constrains the density at
  these scales. While we cannot exclude the presence of a denser
  compact object hidden by the envelope, it seems unlikely that the
  envelope density gradient increases inwards, because this would be
  unphysical.

Clearly, the water abundance in the inner region of FIR4 will have a
great impact not only on the emerging water spectrum but also on the
emerging line spectrum of any molecule (abundant in the inner region),
and has to be correctly taken into account to give reliable molecular
abundances. Conversely, given the large effect, in principle
appropriate multiline observations of any molecule will be able to
constrain the inner region water abundance and the present model
predictions. Note that varying the outer abundance X(H$_2$O)$_{out}$
does not have effect on the gas temperature, as in the outer region
the cooling is dominated by the CO and O lines.

 \section{Predicted water line spectrum}\label{sec:predicted-water-line}
\subsection{Reference model}
Here we report and discuss the predicted spectrum of our reference
model. Next paragraph will discuss how it depends on the parameters of
the model. We adopted the Model 1 of Table \ref{tab:abu-cool} as
reference model . We first discuss the general water line spectrum by
means of the synthetic rotational diagram, and then we discuss the
specific predictions for the two spectrometers on board Herschel: HIFI
and PACS.
\begin{figure} \centering
\rotatebox{0}{\includegraphics[width=9cm]{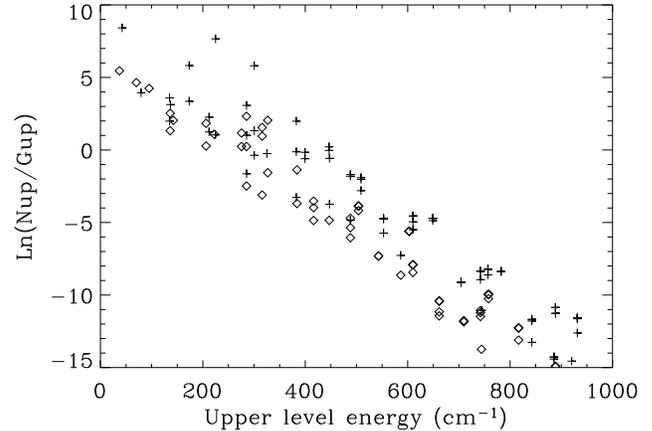}}
\caption{Synthetic rotational diagram derived from the water line emission 
 using the reference model (Model 1, Table \ref{tab:abu-cool})
 integrated over the whole envelope. 
 Crosses and diamonds trace the ortho and para water, respectively.
\label{fig:rot-diagr}}
\end{figure}

\begin{figure} \centering
\rotatebox{0}{\includegraphics[width=9cm]{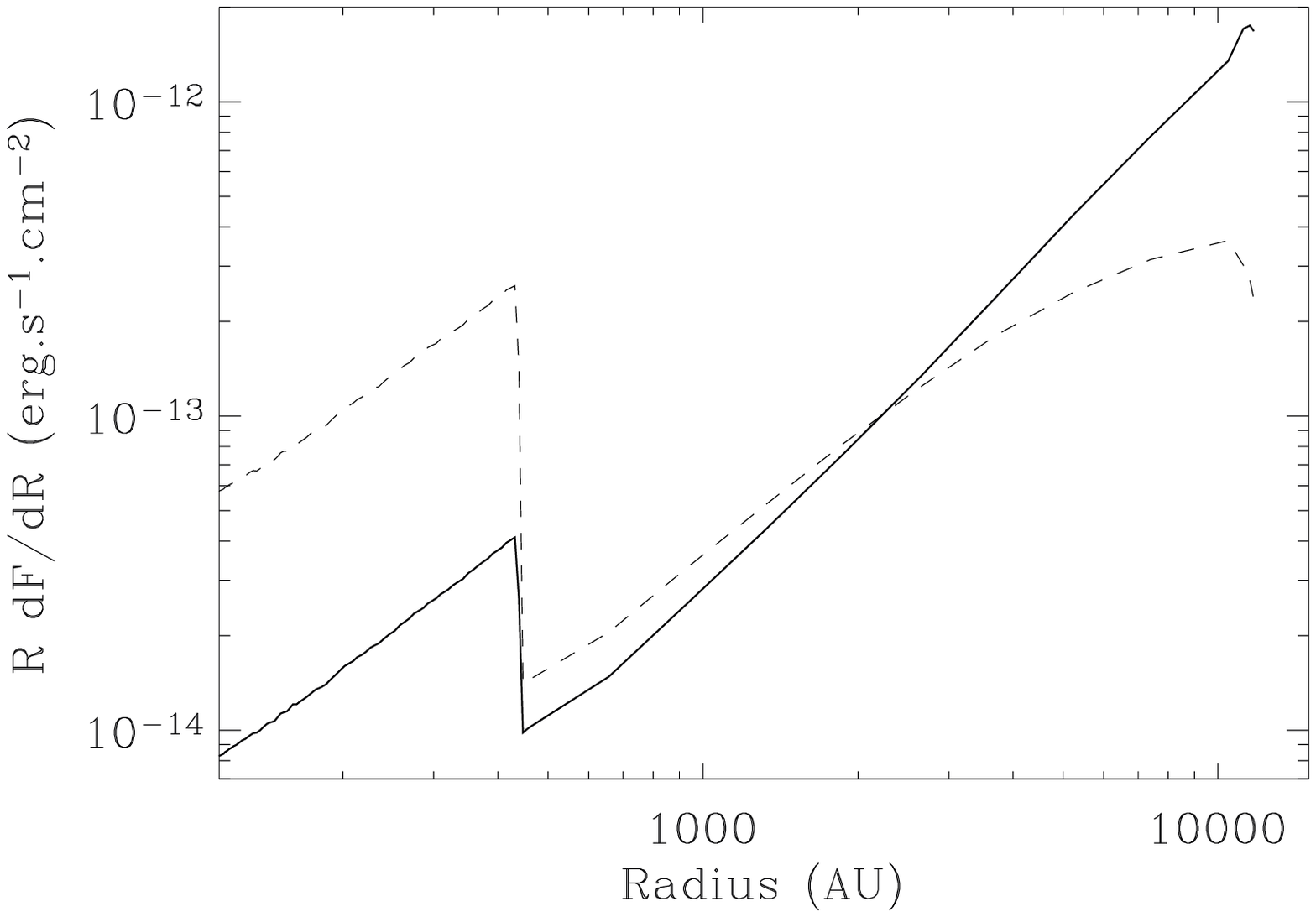}}
\rotatebox{0}{\includegraphics[width=9cm]{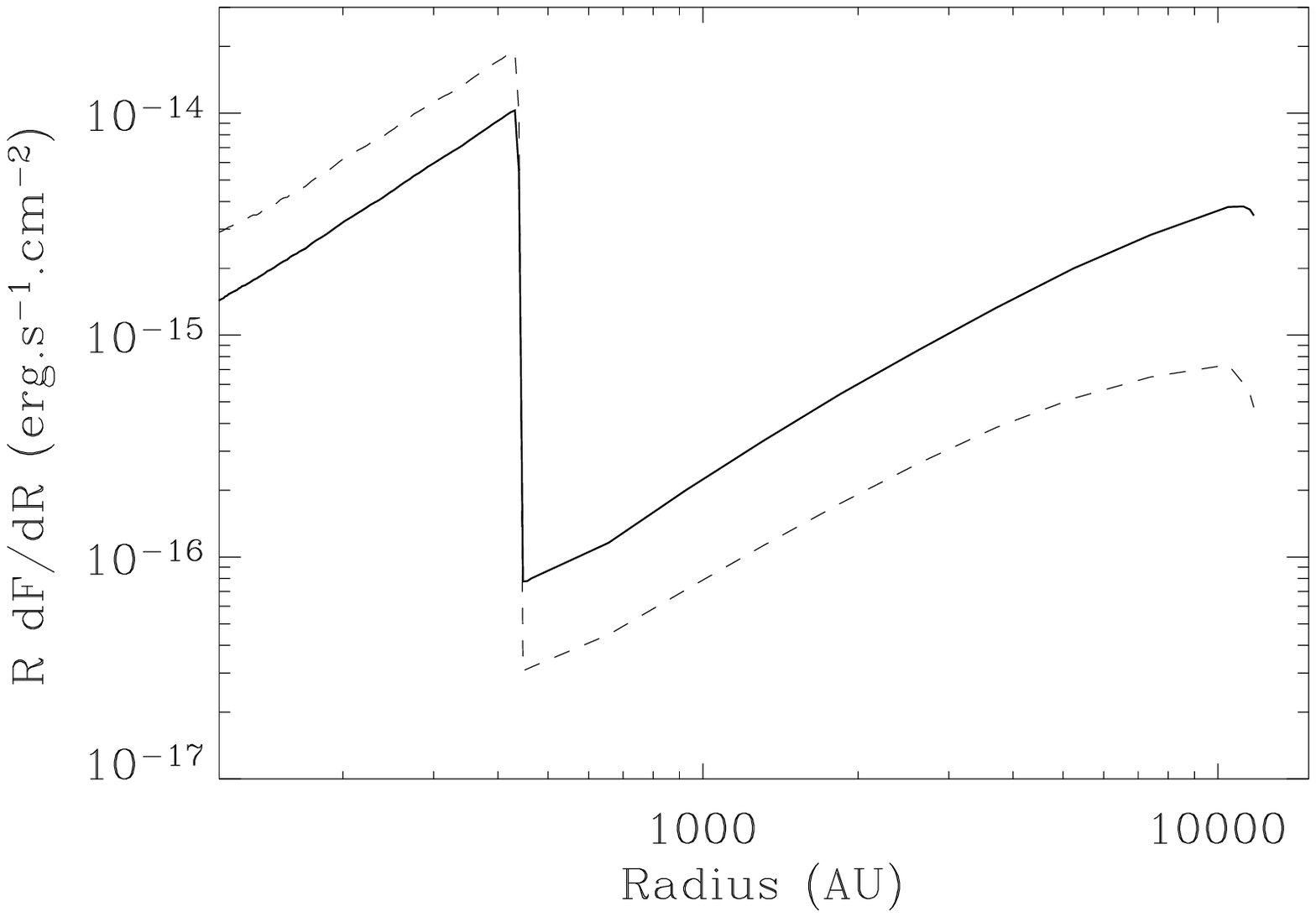}}
\caption[] {Emission profile, $R\frac{dF}{dR}$ of the ortho (solid
  line) and para (dashed line) water fundamental lines at 557 and 1113
  GHz respectively as function of the radius, in the case of the
  reference model (see Table \ref{tab:abu-cool}). The plotted quantity
  $R\frac{dF}{dR}$ is the contribution of each shell to the flux
  integrated over the whole envelope. H$_2^{16}$O and H$_2^{18}$O
  emission profiles are represented on the top and bottom panel,
  respectively.\label{fig:h2o-profile}}
\end{figure}

\begin{figure} \centering
\rotatebox{0}{\includegraphics[width=9cm]{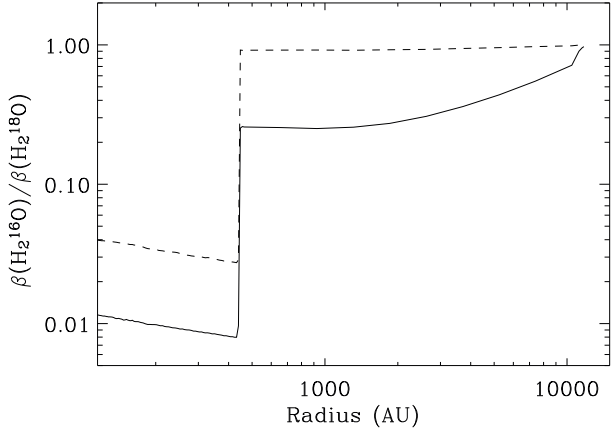}}
\caption[] {Ratio of the H$_2^{16}$O escape probability over the
  H$_2^{18}$O escape probability of the ortho (solid line) and para
  (dashed line) water fundamental lines obtained with the reference
  model (model 1 of Table \ref{tab:abu-cool}) at 557 and 1113 GHz
  respectively as function of the radius.\label{fig:h2o-profile-2}}
\end{figure}

Figure \ref{fig:rot-diagr} shows the synthetic rotational diagram
derived from the line emission integrated over the whole envelope. As
expected, 
 the theoretical points do not lie on a compact and straight
  line, reflecting the different line optical depths, the gradients
in density and temperature of the envelope and non-LTE effects.  An
illuminating example is represented by the fundamental transitions of
the ortho and para water lines at 557 and 1113 GHz respectively. We
will discuss these two lines in detail because, first, they will
certainly be important observational diagnostics and, second, they
offer a great pedagogic case. The situation is illustrated in
Fig.\ref{fig:h2o-profile}, where we report the profile of the emission
of the ortho and para H$_2^{16}$O and H$_2^{18}$O fundamental lines as
function of the radius. Figure \ref{fig:h2o-profile-2}, with the beta
escape probability as function of the radius for the two fundamental
H$_2^{16}$O lines, also greatly helps to interpret the emerging line
fluxes for the two lines.  The ortho-H$_2^{16}$O fundamental line
emission (Fig.\ref{fig:h2o-profile}) peaks at the border of the
envelope and it decreases inwards because of the decreasing emitting
volume.  The para-H$_2$O fundamental line shows approximately the same
behavior.  If the lines were optically thin and LTE populated, the
expected flux ratio of the para over ortho fundamental line would be
between 3 and 4 for a temperature between 50 and 200 K.  Any departure
from this value originates from a combination of line opacity and
non-LTE effects.  In the outer region the ratio is lower than 1: the
para-H$_2$O line is optically thin, whereas the ortho-H$_2$O lines is
moderately optically thick (Fig.\ref{fig:h2o-profile-2}). Therefore,
the much lower emission of the para-H$_2$O line with respect to the
ortho-line is due to the non-LTE population effect, more accentuated
in the para-H$_2$O line.  The situation is reversed in the inner
region, where ices sublimate: the para-H$_2$O fundamental line becomes
about ten times brighter than the ortho-H$_2$O fundamental line
because of the line opacity, which is much larger in the ortho-H$_2$O
line than in the para-H$_2$O line (Fig.\ref{fig:h2o-profile-2}). In
fact, the increase in the water abundance by a factor 1000 gives rise to a jump in the
line emission by a factor 3 in the ortho-H$_2$O line and 30 in the
para-H$_2$O line, and this can only be due to the larger opacity of
the ortho-H$_2$O line as the excitation conditions do not change when
ices sublimate. In summary, the emission from the water lines is due,
in principle, to a rather complex combination of line opacity, non-LTE
effects and emitting volume (namely temperature and density
gradient). Evidently, the intensity ratio of lines from the
H$_2^{16}$O and H$_2^{18}$O isotopologues is far to give the
``opacity'' of the line, as it is a combination of the penetration of
the line and the opacity itself.

 Table \ref{tab:spe-ref} lists the predicted water line fluxes
for the two spectrometers on board Herschel: HIFI and PACS. Note that,
in both cases, we computed the signal after convolving the theoretical line intensity map 
with the instrument beam which vary
  from 39$"$ to 13$"$ with the frequency varying from 500 GHz to 2000
  GHz (HIFI frequency range) and from 13$"$ to 5$"$ for wavelengths
  from 210 $\mu$m and 60 $\mu$m (PACS wavelength range).

\begin{table}[h]
  \centering
  \begin{tabular}{|ccc|}
    \hline
    PACS range & &                                                    \\ \hline
   Transition  & Wavelength & Flux                                    \\
               & ($\mu$m)   & (erg s$^{-1}$ cm$^-2$)                  \\ \hline
  H$_2^{16}$O  &            &                                        \\
  2$_{ 2 1} \rightarrow $ 3$_{ 3 0}$  &       66.44 &  2.99E-14     \\
  2$_{ 2 0} \rightarrow $ 3$_{ 3 1}$  &       67.09 &  1.13E-14     \\
  3$_{ 0 3} \rightarrow $ 3$_{ 3 0}$  &       67.27 &  2.28E-14     \\
  2$_{ 1 2} \rightarrow $ 3$_{ 2 1}$  &       75.38 &  1.55E-13     \\
  3$_{ 1 2} \rightarrow $ 4$_{ 2 3}$  &       78.74 &  4.51E-14     \\
  2$_{ 1 1} \rightarrow $ 3$_{ 2 2}$  &       89.99 &  5.09E-14     \\
  4$_{ 0 4} \rightarrow $ 5$_{ 1 5}$  &       95.63 &  1.05E-14     \\
  4$_{ 1 4} \rightarrow $ 5$_{ 0 5}$  &       99.49 &  4.21E-14     \\ 
  1$_{ 1 1} \rightarrow $ 2$_{ 2 0}$  &      100.98 &  7.80E-14     \\ 
  1$_{ 1 0} \rightarrow $ 2$_{ 2 1}$  &      108.07 &  1.43E-13     \\
  3$_{ 0 3} \rightarrow $ 4$_{ 1 4}$  &      113.54 &  8.45E-14     \\
  4$_{ 2 3} \rightarrow $ 4$_{ 3 2}$  &      121.72 &  1.14E-14     \\
  3$_{ 1 3} \rightarrow $ 4$_{ 0 4}$  &      125.36 &  5.34E-14     \\
  4$_{ 1 4} \rightarrow $ 4$_{ 2 3}$  &      132.41 &  2.92E-14     \\
  3$_{ 2 1} \rightarrow $ 3$_{ 3 0}$  &      136.49 &  1.35E-14     \\
  2$_{ 0 2} \rightarrow $ 3$_{ 1 3}$  &      138.53 &  7.84E-14     \\
  3$_{ 1 3} \rightarrow $ 3$_{ 2 2}$  &      156.20 &  2.25E-14     \\
  2$_{ 1 2} \rightarrow $ 3$_{ 0 3}$  &      174.62 &  7.42E-14     \\
  1$_{ 0 1} \rightarrow $ 2$_{ 1 2}$  &      179.53 &  1.17E-13     \\
  2$_{ 1 2} \rightarrow $ 2$_{ 2 1}$  &      180.49 &  4.97E-14     \\ \hline
HIFI range & & \\ \hline
   Transition  & Frequency  & Flux \\
               & (GHz)      & (K Km s$^{-1}$) \\ \hline
 H$_2^{16}$O  &            &                 \\
  1$_{ 0 1} \rightarrow $ 1$_{ 1 0}$  &      556.96 &  14.8 \\
  2$_{ 0 2} \rightarrow $ 2$_{ 1 1}$  &      752.04 &  1.06E+00 \\
  1$_{ 1 1} \rightarrow $ 2$_{ 0 2}$  &      987.95 &  1.31E+00 \\
  3$_{ 0 3} \rightarrow $ 3$_{ 1 2}$  &     1097.34 &  1.46E+00 \\
  0$_{ 0 0} \rightarrow $ 1$_{ 1 1}$  &     1113.35 &  4.32E+00 \\
  2$_{ 2 1} \rightarrow $ 3$_{ 1 2}$  &     1153.09 &  2.50E+00 \\
  3$_{ 1 2} \rightarrow $ 3$_{ 2 1}$  &     1162.93 &  6.97E-01 \\
  2$_{ 1 1} \rightarrow $ 2$_{ 2 0}$  &     1228.81 &  5.33E-01 \\ 
  4$_{ 1 3} \rightarrow $ 4$_{ 2 2}$  &     1207.62 &  2.45E-01 \\ 
  5$_{ 1 4} \rightarrow $ 5$_{ 2 3}$  &     1410.65 &  1.47E-01 \\
  2$_{ 1 2} \rightarrow $ 2$_{ 2 1}$  &     1660.99 &  2.33E+00 \\
  4$_{ 0 4} \rightarrow $ 4$_{ 1 3}$  &     1602.23 &  1.46E-01 \\
  1$_{ 0 1} \rightarrow $ 2$_{ 1 2}$  &     1669.87 &  5.46E+00 \\
  2$_{ 1 2} \rightarrow $ 3$_{ 0 3}$  &     1716.83 &  3.37E+00 \\
  5$_{ 2 3} \rightarrow $ 5$_{ 3 2}$  &     1867.75 &  3.13E-02 \\
 H$_2^{18}$O  &            &                 \\
  1$_{ 0 1} \rightarrow $ 1$_{ 1 0}$  &      556.96 &  1.88E-01 \\
                 &            &                                 \\
  0$_{ 0 0} \rightarrow $ 1$_{ 1 1}$  &     1113.35 &  1.48E-01 \\
 &  &\\ \hline
  \end{tabular}
  \caption{Predictions of the line fluxes (after subtraction of the continuum) 
    of the water lines observable with the Herschel spectrometers, HIFI and 
    PACS. The predictions refer to the reference model (model 1 of Table 
    \ref{tab:abu-cool}). \label{tab:spe-ref}}
\end{table}

 Based on the (preliminary) sensitivities reported on the Herschel
  Observation Planning Tool HSpot: {\it
    http://herschel.esac.esa.int/Tools.shtml}), several ortho and
para lines are predicted to be detectable by the two Herschel
spectrometers: about a dozen in the HIFI frequency range and twice
more in the PACS wavelength range. The H$_2^{18}$O ortho and para
lines are also predicted to be detectable by HIFI, and 100 and 20
times less bright than the respective lines of the H$_2^{16}$O
respectively. Note the counter-intuitive result: the para-H$_2^{16}$O
line seems to be more optically thick than the ortho-H$_2^{16}$O line!
As explained above, this is not the case, of course: the line
intensity ratio (from which the line optical depth is usually derived) is due to the combination of optical depth plus
excitation (non-LTE) effects, and the final result is not easily predictable. In our
reference model, no observable line is predicted to be in
absorption.

\subsection{Other models}
Here we explore the sensitivity of the results reported in the
previous paragraph against the variation of the three main parameters
of the model: the water abundance in the inner (X(H$_2$O)$_{in}$) and
outer (X(H$_2$O)$_{out}$) envelope, and the illuminating FUV field
G$_o$. 

\begin{figure} \centering
\rotatebox{90}{\includegraphics[width=6cm]{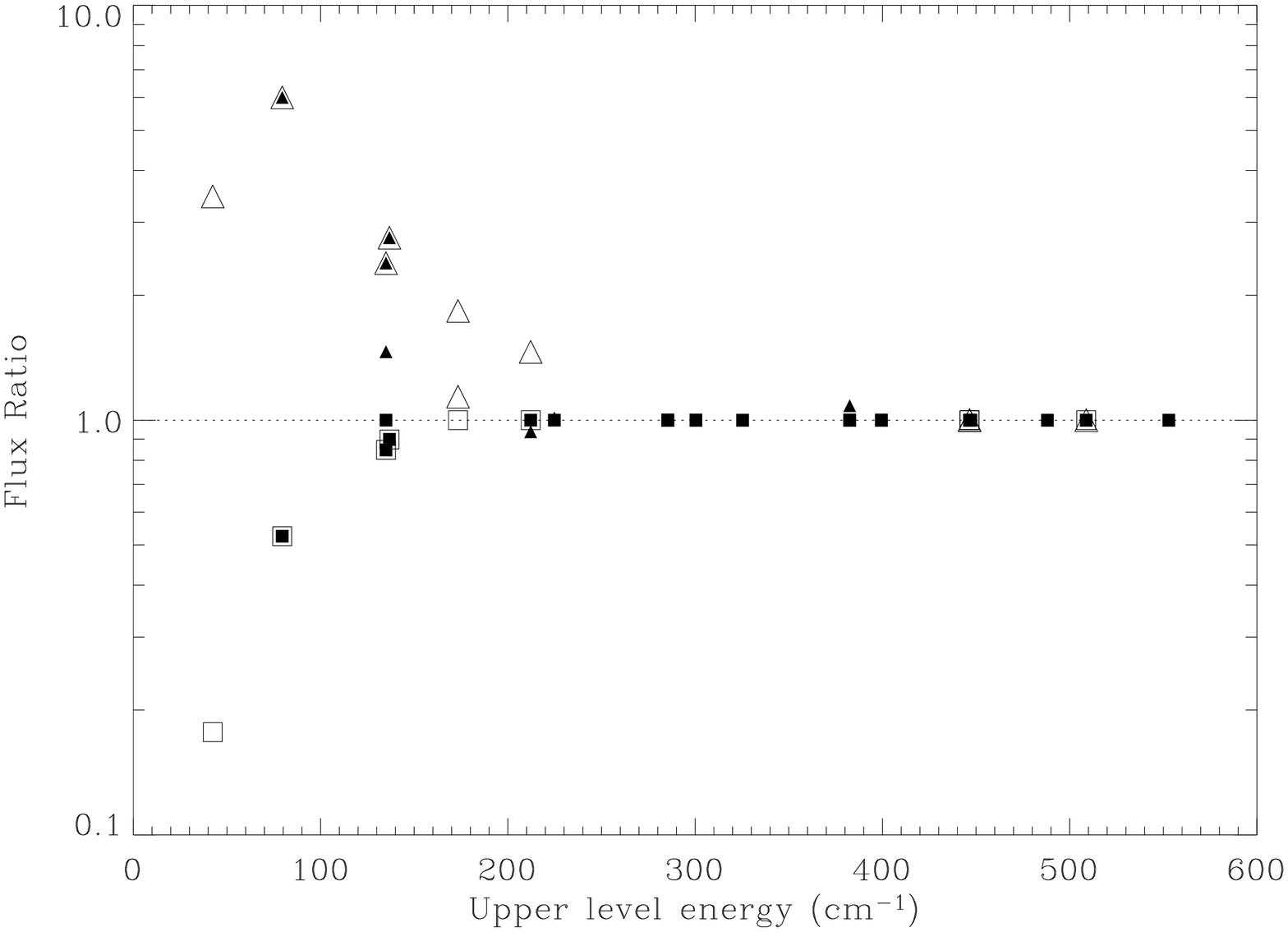}}
\rotatebox{90}{\includegraphics[width=6cm]{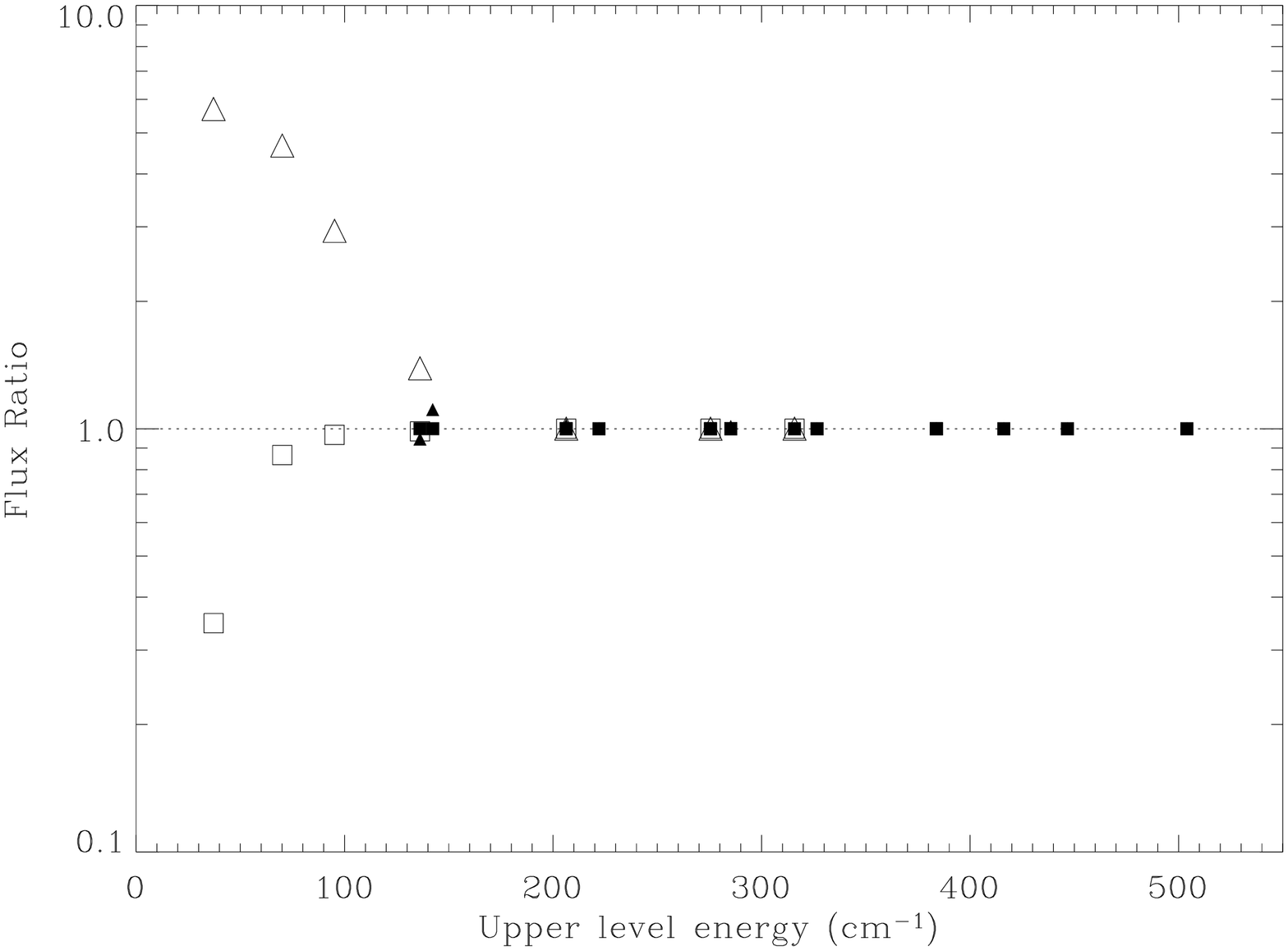}}
\caption[] {Ratio between the line fluxes of the reference model
  (model 1 of Table \ref{tab:abu-cool}) and the line fluxes predicted
  by models with same X(H$_2$O)$_{in}$ but different
  X(H$_2$O)$_{out}$. Flux ratios between model 4 (model 5) and model 1
  are represented by squares (triangles). Filled and empty symbols
  refer to the lines emitted in the PACS and HIFI bands,
  respectively. The upper (lower) panel reports ortho (para) water
  line intensities ratios. \label{hifi_ratio_xout}}
\end{figure}

\begin{figure} \centering
\rotatebox{90}{\includegraphics[width=6cm]{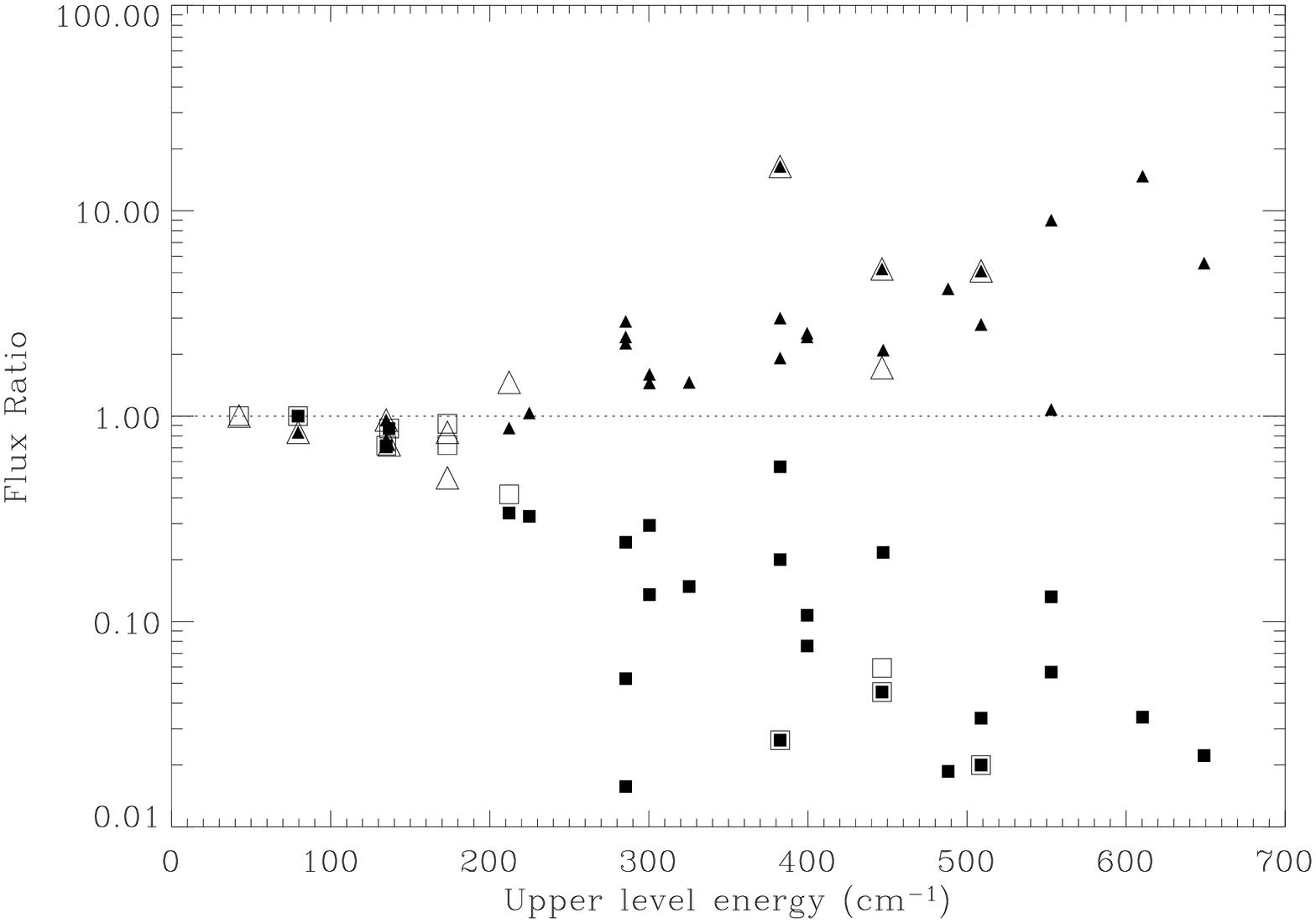}}
\rotatebox{90}{\includegraphics[width=6cm]{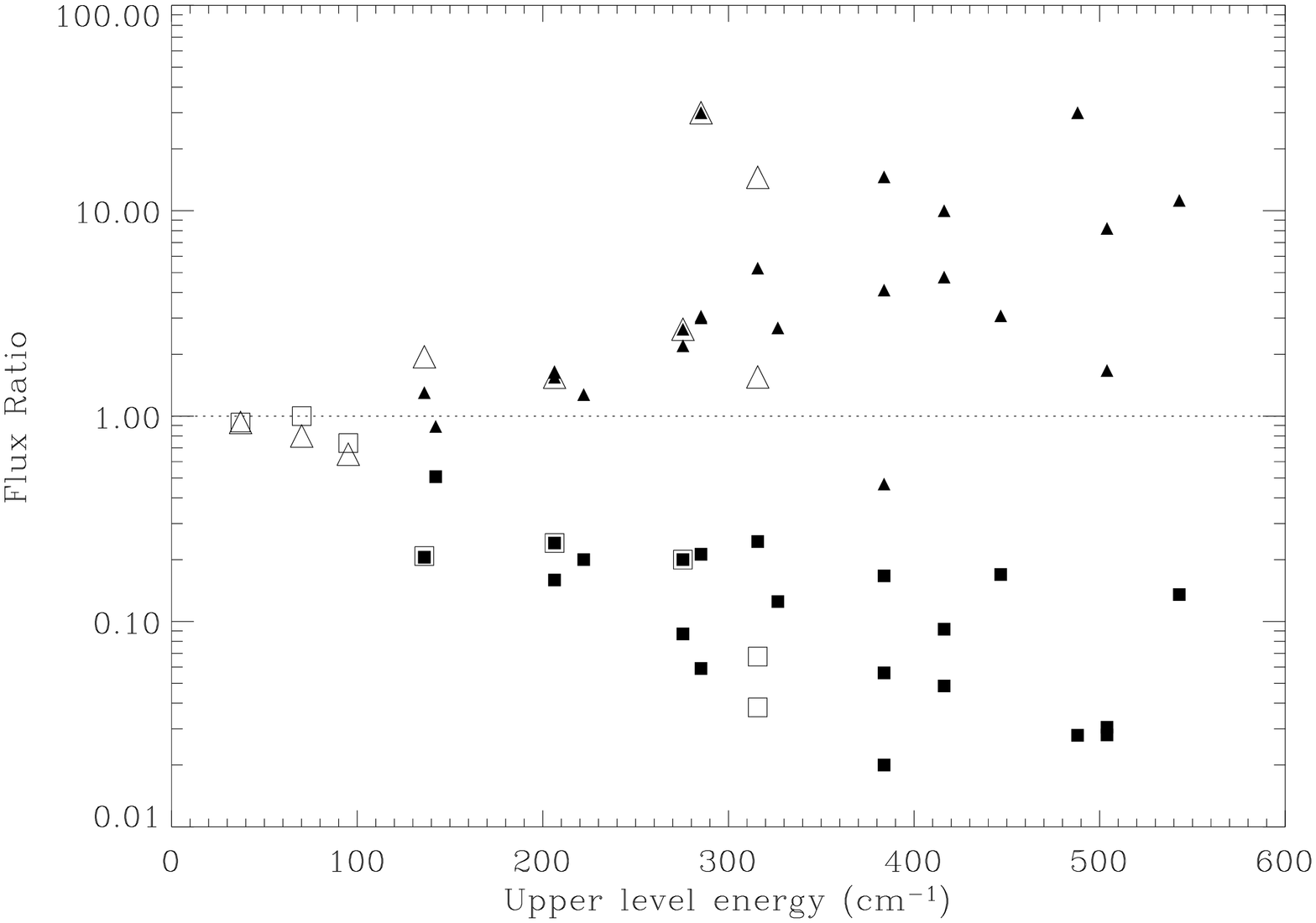}}
\caption[] {Ratio between the line fluxes of the reference model
  (model 1 of Table \ref{tab:abu-cool}) and the line fluxes predicted
  by models with same X(H$_2$O)$_{out}$ but different
  X(H$_2$O)$_{in}$. Flux ratios between model 2 (model 3) and model 1
  are represented by squares (triangles). Filled and empty symbols
  refer to the lines emitted in the PACS and HIFI bands,
  respectively. The upper (lower) panel reports ortho (para) water
  line flux ratios. \label{hifi_ratio_xin}}
\end{figure}

Figures \ref{hifi_ratio_xout} and \ref{hifi_ratio_xin} show the ratio
between the line intensities of the reference model (Model 1 of Table
\ref{tab:abu-cool}) and the line intensities predicted by models with
different X(H$_2$O)$_{out}$ and X(H$_2$O)$_{in}$ respectively. As
noted by other authors \citep{Cec00,Mar02}, lines with upper level
energies lower than about 200 cm$^{-1}$ are sensitive to
X(H$_2$O)$_{out}$ and insensitive to X(H$_2$O)$_{in}$, because these
lines mostly originate in the outer envelope for excitation and line
opacity reasons. A variation of a factor 10 in X(H$_2$O)$_{out}$ leads
to an almost similar variation in the line intensity of the lowest
lying lines. The higher the upper level energy the smaller the
variation.  Conversely, lines with upper level energies larger than
about 200 cm$^{-1}$ are sensitive to X(H$_2$O)$_{in}$ and insensitive
to X(H$_2$O)$_{out}$. In this case, variations by a factor 10 in
X(H$_2$O)$_{in}$, going from $1\times10^{-6}$ to $1\times10^{-5}$, can
lead to variations in the lines fluxes even 100 times larger. This
extreme variation, 10 times larger than the difference in the
X(H$_2$O)$_{in}$ variation, occurs to some lines in the 50-200 $\mu$m
wavelength range. This phenomenon is due to the fact that those lines
are in absorption rather than in emission in the region just after the
ices sublimation, resulting in an additional decrease of the emerging
line flux. The higher the X(H$_2$O)$_{in}$ the smaller the absorption
depth. When X(H$_2$O)$_{in}$ reaches $1\times10^{-5}$ the absorption
region generally vanishes. In addition, many high lying lines are
  prevalently populated by absorption of the photons emitted by the
  dust, so that they are particularly sensitive to the dust
  continuum.

Note that as mentioned in the Sect. \ref{sec:cont-emiss-dusty}, the inner region of the envelope is relatively unconstrained by the available observational data. Therefore we derived water line spectrum predictions varying the power law index of the density profile $\alpha$ of about $\sim$ 30$\%$ in the inner part. We observed a variation of the line intensity of a factor 5 - 10 for the transitions with upper level energy $\gtrsim$ 300$-$400 cm$^{-1}$ and lesser than 2 for the lower lines.
Finally, the predicted line intensities do not vary appreciably when
the illuminating FUV field changes from 1 to 1000. Therefore, observations of 
water lines will be extremely helpful in constraining the water abundance across
the envelope, but will not be sensitive to the illuminating FUV field.

\subsection{Effect of gas-dust thermal decoupling}

\begin{figure} \centering
\rotatebox{90}{\includegraphics[width=6cm]{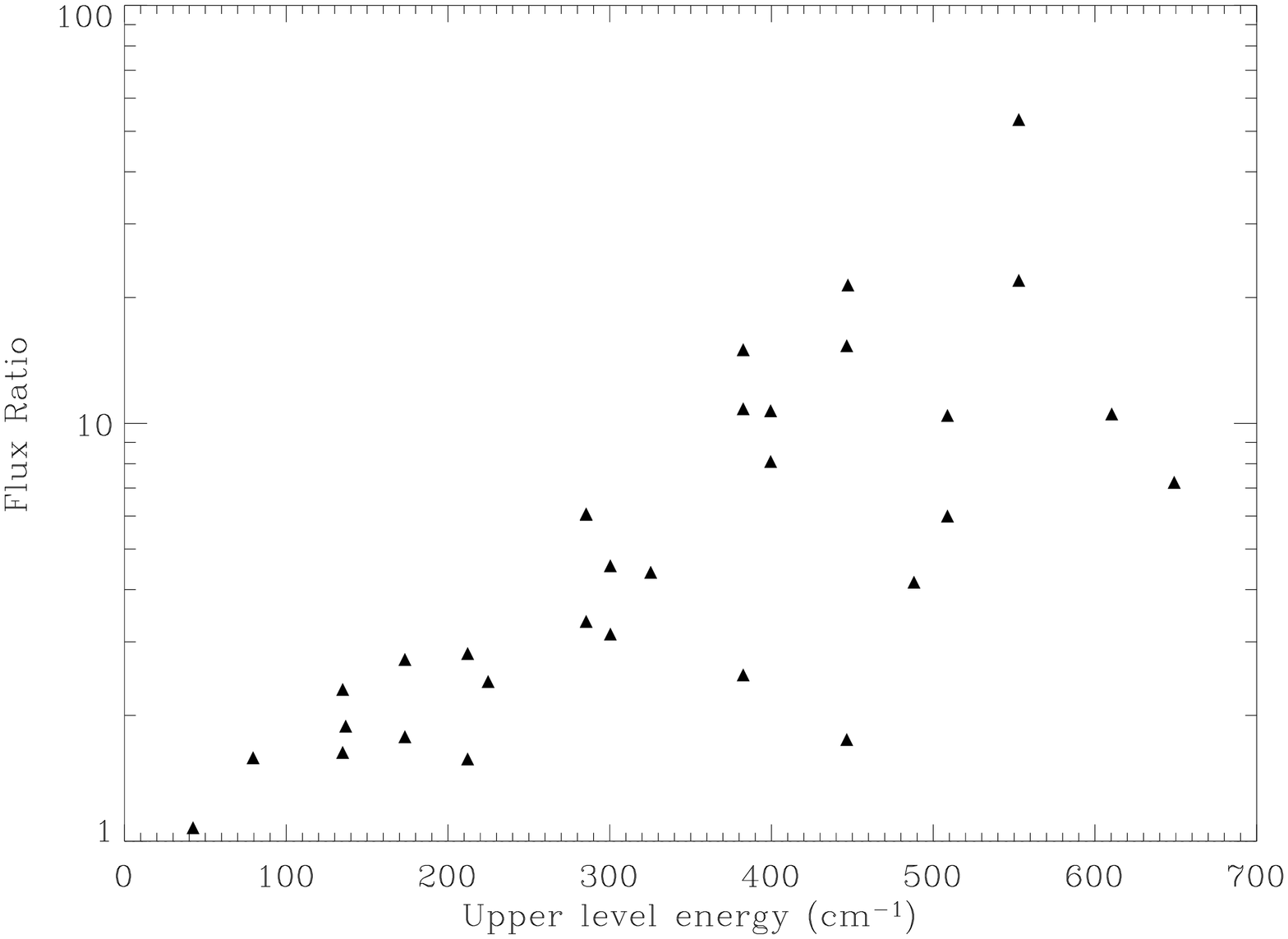}}
\rotatebox{90}{\includegraphics[width=6cm]{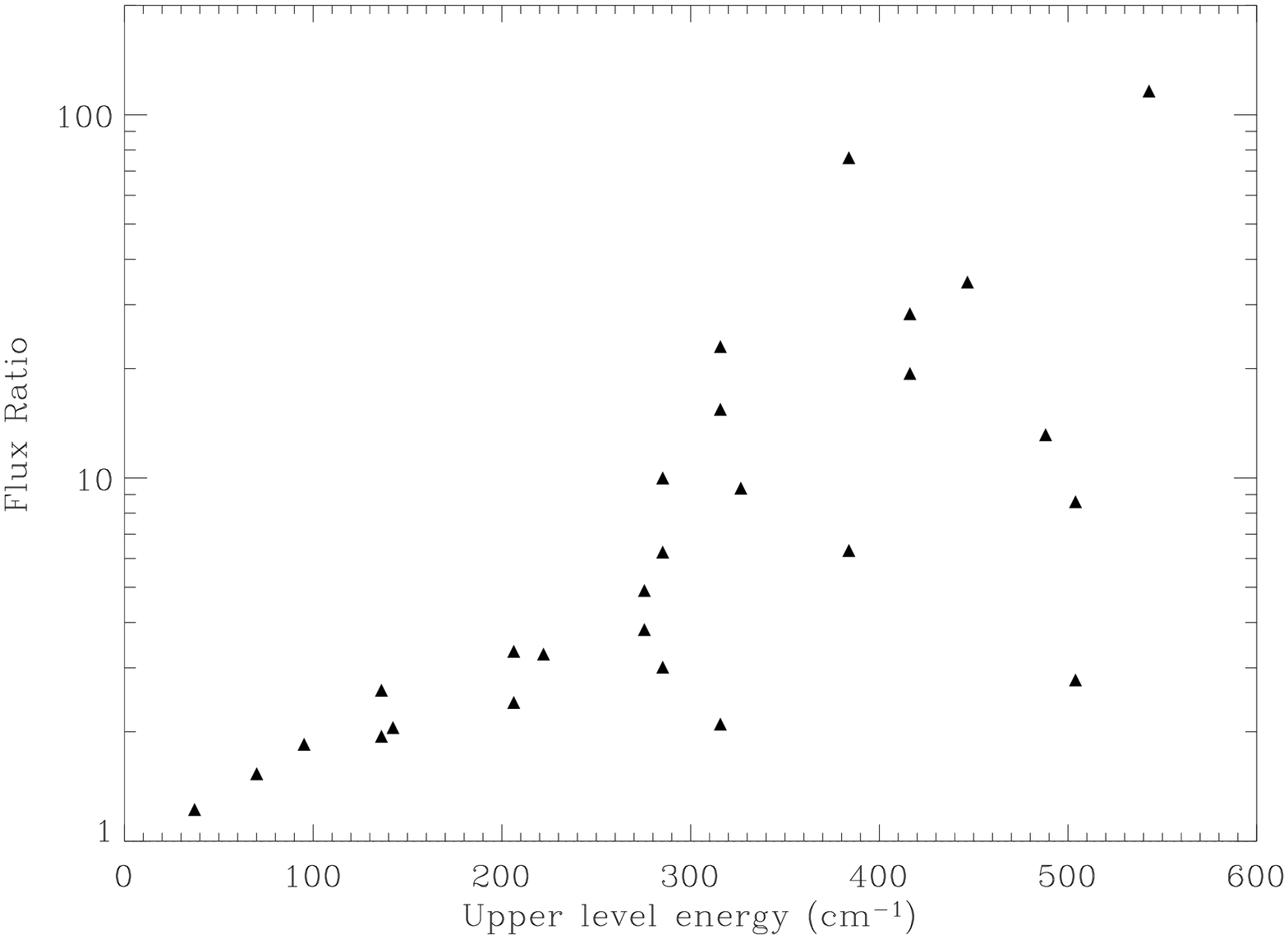}}
\caption[]{Ratios between the line fluxes of the reference model
  (model 1) and the reference model with gas-dust non-thermally
  decoupled (namely T$_{gas}$=T$_{dust}$, as function of the upper
  level energy of the transition. Upper and lower panels show ortho
  and para H$_2$O lines, respectively. \label{ratio_decoup}}
\end{figure}

As presented in \S \ref{sec:gas-temp-prof}, the large quantity of
water vapor injected into the gas in the inner part of the envelope
causes a dramatic decoupling between the dust and gas temperatures
(see Fig. \ref{tg.ps}). Obviously, this effect has a great impact in
the interpretation of the water line emission. This is illustrated in
Figures \ref{ratio_decoup}, where we report the ratio of the water
line intensities obtained by considering the gas temperature
self-consistently computed (model 1) over the case where T$_{gas}$ is
assumed to be equal to T$_{dust}$.
Assuming artificially T$_{gas}$ = T$_{dust}$ leads to differences in
line fluxes up to two orders of magnitude.  Since the decoupling
occurs in the inner part of the envelope, the larger the upper level
energy the larger the difference, for both ortho and para lines. Note
that the fluxes of the two fundamental ortho and para lines are not
affected by the T$_{gas}$ = T$_{dust}$ choice. Finally, we did the
same study with the model 2 ( X(H$_2$O)$_{in}$ = 1O$^{-6}$). In this
case, due to the smaller X(H$_2$O)$_{in}$, the decoupling is less
important than in the previous one, leading to changes in lines
intensities up to one order of magnitude.

\subsection{Constraints from ISO data}\label{sec:iso-data:water}
Observations of FIR4 were obtained by the spectrometer ISO-LWS in the
grating mode (spectral resolution $\sim$200) and Fabry-Perot (spectral
resolution $\sim 10^{4}$). We retrieved the data from the ISO Data
Archive ({\it http://iso.esac.esa.int/ida/}). No water lines are
detected. Table \ref{ISO_upper} summarizes the upper limits obtained
for the predicted brightest lines together with the predictions of the
reference model and model 6.
\begin{table*}
 \centering
\begin{tabular}{|cc|c|c|cc|} \hline 
&&Model 1 &Model 6          & ISO & \\
Wavelength    &  Transition        &   Intensity &   Intensity   & Instrument  & Upper Limit (3$\sigma$)  \\
 ($\mu$m)     & J$_{K_- K_+} \rightarrow$ J'$_{K'_- K'_+}$ &   (erg.s$^{-1}$.cm$^{-2}$) &   (erg.s$^{-1}$.cm$^{-2}$) &&  (erg.s$^{-1}$.cm$^{-2}$) \\ \hline
ortho &&&&& \\
75.4957       & 8$_{5 4} \rightarrow$  8$_{4 5}$       &  2.6E-20        &      1.9E-19         & LWS04       & 9.78e-12 \\
108.073       & 2$_{2 1} \rightarrow$  1$_{1 0}$       &  1.3E-12        &      1.5E-12         & LWS04       & 2.21e-12 \\ 
179.527       & 2$_{1 2} \rightarrow$  1$_{0 1}$       &  1.2E-12        &      7.0E-12         & LWS04       & 2.56e-11 \\  
113.538       & 4$_{1 4} \rightarrow$  3$_{0 3}$       &  8.3E-13        &      8.7E-13         & LWS01       & 1.2e-11 \\  
para &&&&& \\
100.983       & 2$_{2 0} \rightarrow$  1$_{1 1}$       &  7.3E-13        &      9.1E-13         & LWS01       & 8.37e-12 \\
138.527       & 3$_{1 3} \rightarrow$  2$_{0 2}$       &  7.3E-13        &      7.3E-13         & LWS01       & 4.14e-12 \\
156.197       & 3$_{2 2} \rightarrow$  3$_{1 3}$       &  2.2E-13        &      3.4E-13         & LWS01       & 1.60e-12 \\ \hline
\end{tabular}
\caption{The brightest lines predicted by models 1 and 6 (the model with the largest water abundance) compared with the upper limits derived by the ISO observations.)  \label{ISO_upper}}
\end{table*}
Unfortunately, the ISO sensitivity is not enough to put sensible
constraints to the water abundance across the FIR4 envelope.


\section{Concluding remarks}\label{sec:conclusion}
We have analyzed in great detail the continuum emission from the
Intermediate Mass protostar OMC2-FIR4, with the aim of deriving the
physical structure of its envelope, a mandatory first step for further
studies to understand the formation process. Our analysis led to a new
estimate of the FIR4 luminosity, which is 1000 L$_\odot$. The density
of the envelope surrounding FIR4 has a shallow dependence on the
radius, the density power law index being only 0.6. Since
  systematic studies of the IM protostars envelopes have not been
  published yet, we can tentatively compare the FIR4 envelope with low
  and high mass protostellar envelopes, where similar studies have been
  carried out. Specifically, \citet{Jor02} analyzed 18 Class 0 and I
  sources and found that the average power law index $\alpha$ in Class
  0 sources is $1.3\pm0.4$ while in Class I sources it is $1.7\pm0.1$,
  significatively larger than the value we found in FIR4. Similarly,
  \citet{Van00} studied a sample of high mass protostars
  and found $\alpha=1.4\pm0.4$. One has to notice that, however, there
  are exceptions, with sources in both low and high mass showing a
  smaller than unity $\alpha$ value: L1527 ($\alpha=0.6$) and L483
  ($\alpha=0.9$) in the Class 0 sources, GL7009S ($\alpha=0.5$) in the
  high mass protostars sample. It is not clear what makes these
  sources ``anomalous'': the presence of strong asymmetries \citep{Jor02} have been suggested as possible reason. The
  case of FIR4 seems to fall in this ``anomalous sources'' category,
  and further studies, possibly on chemistry, are needed to say more.

  Giving the suggestion by \citet{Jor06} that a strong FUV field
  (G$_0$=$1\times10^{4}$) illuminates the FIR4 envelope, we explored
  the cases of different FUV fields.  As already noted by the same
  authors, however, the dust continuum cannot really distinguish
  whether a strong illuminating FUV field is impinging on the
  envelope. In fact, \citet{Jor06} adopted a steeper density distribution ($\alpha$=2) which allows the FUV photons to penetrate deeper into the envelope. Their conclusions were based on
  submillimeter lines from CO and H$_2$CO, which would be exceedingly
  bright if they were emitted in the envelope. They attributed the
  lines to the warm gas at the border of the envelope, heated up by
  the hypothetical large FUV field. However, as discussed in \S
  \ref{sec:dust-results}, OI and CII maps by \citet{Her97} showed
  that the entire OMC2 region is illuminated by a G$_0$=500 FUV field,
  which would imply an even lower FUV field on the FIR4 envelope. One
  has also to notice here that large scale maps by \citet{sch82} show
  that the CO (1$\rightarrow$0) line is bright ($\sim$40 K) over the
  whole OMC2 region, a fact that lead \citet{Her97} to attribute the
  CO emission to the PDR associated with the cloud. In addition,
  several outflows are known to "pollute" the CO emission in the
  region, in particular the one originating from FIR3 (25$"$
  North of FIR4: Fig. \ref{maps}) and reaching FIR4 and FIR5
  \citep{Wil03}.  
 All the above considerations together lead to
conclude that the FUV field impinging FIR4 is not anomalously large
and less than 500. Therefore, giving the presence of a bright PDR
  and a ``polluting'' outflow from FIR3, caution is needed in
  interpreting the low lying water lines, as much as lines from any
  molecule, separately from the whole molecular cloud emission.

One major motivation of the present work is the prediction of the
water line spectrum from FIR4, as this source will be observed in the
500-2000 GHz frequency range by the incoming Herschel Space
Observatory (FIR4 is a target of the Key Program ``HIFI Spectral
Surveys of Star Formation Regions'': {\it
  http://www-laog.obs.ujf-grenoble.fr/heberges/hs3f/}). In the present
study, we have shown that water is indeed a key molecular species,
because of its great impact on the gas cooling in the region where the
dust temperature exceeds 100 K, the sublimation temperature of the
dust grain ices. The large quantity of water vapor injected into the
gas by the sublimated ices very efficiently cools the gas, causing a
dramatic decoupling between the dust and gas temperatures. Depending
on the abundance of the injected water vapor, the difference in the
temperature can be as high as 50 K at the sublimation radius (namely
50\%!) and even larger going inward. For example, at 100 AU the dust
temperature is predicted to be around 280 K whereas the gas
temperature is 80 K if the water abundance is
$1\times10^{-4}$. Obviously, this has a great impact in the
interpretation of the water line emission as much as the emission from
any molecular species emitting in the inner region. In fact, the
comparison of the water line emission between the case where dust and
gas are assumed to be thermally coupled and the case where the gas
temperature is self-consistently computed shows that the difference of
the line intensity can reach two orders of magnitudes for lines with
large upper level energies (namely the lines excited in the innermost
region, where gas and dust decouple). Therefore, our important second
conclusion is that caution has to be applied in interpreting the line
emission from FIR4, as much as any source with a similar luminosity
and envelope structure. Gas and dust temperature can be very different
and in order to derive correct molecular abundances (including water
abundance) they have to be both estimated, accounting for all terms of
heating and cooling. Avoiding that may lead to very wrong conclusions.

\bigskip
\begin{acknowledgements}
  We warmly thank Moshe Elitzur for his valuable help in using the
  DUSTY code. We also wish to thank Neal Evans and Doug Johnstone for
  helpful discussions, and Doug Johnstone and Darek Lis for providing
  us with the JCMT and CSO continuum maps of the OMC2-FIR4 region. We thank an anonymous referee and Malcolm Walmsley for comments which helped improving the manuscriptOne of us (N.Crimier) is supported by a fellowship of the Minist\`ere de
l'Enseignement Sup\'erieur et de la Recherche. 
\end{acknowledgements}

\bibliographystyle{aa}
\bibliography{bib}

\end{document}